\documentclass[11pt]{article}
\pdfoutput=1

\usepackage{graphics, color,soul}
\usepackage{graphicx}
\usepackage{amssymb}

\usepackage{lmodern,mathtools}

\usepackage{booktabs}
\usepackage[english]{babel}
\usepackage{amsmath,amssymb,amsbsy,amstext, amsthm, simplewick}
\usepackage{hyperref}
\usepackage{tikz}

\usetikzlibrary{decorations.pathmorphing,shapes.misc}
\tikzset{snake it/.style={decorate, decoration=snake}}
\tikzset{cross/.style={cross out, draw=black, minimum size=2*(#1-\pgflinewidth), inner sep=0pt, outer sep=0pt},
cross/.default={1pt}}

\usepackage{amsfonts}
\usepackage{amssymb}
\usepackage{upgreek}
\usepackage{simplewick}
 \usepackage{exscale,relsize}
\usepackage{mathtools}
\usepackage{comment}

\usepackage{subfig}

\usepackage[labelfont={bf}]{caption}

\usepackage{colortbl}
\definecolor{lightgreen}{cmyk}{0.2, 0, 0.2, 0.2}
\definecolor{lightgray}{cmyk}{0.1,0.2,0,0.1}
\definecolor{lightgray2}{cmyk}{0.1,0.1,0,0.1}

\makeatletter
\newlength{\apb@width}
\newcommand{\autoparbox}[2][c]{\settowidth{\apb@width}{#2}\parbox[#1]{\apb@width}{#2}}

\makeatother

\numberwithin{equation}{section}

\def\beq{\begin{equation}}
\def\eeq{\end{equation}}

\def\bea{\begin{eqnarray}}
\def\eea{\end{eqnarray}}

\def\D{\nabla}

\def\beq{\begin{equation}}
\def\eeq{\end{equation}}
\def\be{\begin{equation}}
\def\ee{\end{equation}}
\def\bea{\begin{eqnarray}}
\def\eea{\end{eqnarray}}

\def\C{{\hat{C}}}
\def\D{{\hat{D}}}

\def\0{{\vec{0}}}
\def\q{{\vec{q}}}
\def\p{{\vec{p}}}

\def\p{{\bf p}}

\DeclareRobustCommand{\SkipTocEntry}[4]{}

\def\beq{\begin{equation}}
\def\eeq{\end{equation}}

\def\ba#1\ea{\begin{align}#1\end{align}}
\def\bg#1\eg{\begin{gather}#1\end{gather}}
\newcommand{\bseq}{\begin{subequations}}
\newcommand{\eseq}{\end{subequations}}

\renewcommand{\Im}{\mathop{\rm Im}\nolimits}
\renewcommand{\Re}{\mathop{\rm Re}\nolimits}

\DeclareSymbolFont{extraup}{U}{zavm}{m}{n}
\DeclareMathSymbol{\varheart}{\mathalpha}{extraup}{86}
\DeclareMathSymbol{\vardiamond}{\mathalpha}{extraup}{87}

\def\({\left(}
\def\){\right)}
\def\[{\left[}
\def\]{\right]}

\begin{document}

\begin{titlepage}

\setcounter{page}{1} \baselineskip=15.5pt \thispagestyle{empty}

\vbox{\baselineskip14pt
%%%\hbox{hep-th/0000000}
}
{~~~~~~~~~~~~~~~~~~~~~~~~~~~~~~~~~~~~
~~~~~~~~~~~~~~~~~~~~~~~~~~~~~~~~~~
~~~~~~~~~~~ }

\bigskip\

\vspace{2cm}
\begin{center}
{\fontsize{23}{36}
\selectfont
Recovering Infalling Information via String Spreading
}
\end{center}

\vspace{0.6cm}

\begin{center}
{\fontsize{13}{30}\selectfont  Alexandros Mousatov and Eva Silverstein}
\end{center}

\vspace{0.2cm}

\begin{center}
\vskip 8pt

\textsl{
\emph{$^1$Stanford Institute for Theoretical Physics, Stanford University, Stanford, CA 94306}}

%\vskip 7pt
%\textsl{ \emph{$^2$Kavli Institute for Theoretical Physics, University of California, Santa Barbara, CA 93106}}

%\vskip 7pt
%\textsl{ \emph{$^2$SLAC National Accelerator Laboratory, 2575 Sand Hill, Menlo Park, CA 94025}}

%\vskip 7pt
%\textsl{ \emph{$^2$Kavli Institute for Particle Astrophysics and Cosmology, Stanford, CA 94025}}

\end{center}

\vspace{0.5cm}
\hrule \vspace{0.1cm}
{ \noindent \textbf{Abstract}   

We find S-matrix evidence that longitudinal string spreading can induce interactions between early and late time systems in the near-horizon region of a black hole. By generalizing the effect to closed strings and performing an eikonal resummation, we find a tractable regime where these interactions become strong at a Schwarzschild time separation $\Delta t \gg r_s$.
We estimate the mutual information in a scenario analogous to Hayden-Preskill, and we find that string spreading is sufficient in itself for a late-time detector to recover a significant fraction of the information encoded in an infaller's state.  Interesting open directions include analysis of the interaction of the detector with Unruh radiation (which may introduce noise that somewhat degrades the recovery), and formulating the optimal detector setup including many entangled detectors (which could further increase the information recovery by enhancing sensitivity to kinematic parameters with subleading dependence in the amplitude).

\vspace{0.3cm}

\hrule
\vspace{0.6cm}}
\end{titlepage}

\tableofcontents

\section{Introduction}

The aim of this paper is to show that the longitudinal spreading of strings \cite{lennyspreading, BHpaper}, recently corroborated by open string S-matrix calculations in \cite{sixpoints, dilatontracer}, provides the required nonlocality for a nontrivial fraction of an infaller's information to escape the black hole, while still remaining under perturbative control.

Once we complete our main analysis of the interaction between an early infaller and late detector, we will comment briefly on open questions regarding additional effects that may affect the information recovery.  This includes the possibility of the detector being perturbed by Unruh radiation, which introduces noise at some level.  Going the other way,  we raise the prospect of a more optimal detector setup than the simple one that we will work with here that could increase its statistical sensitivity 
to the kinematical parameters of the early infaller.   

A string $C$ quantized with respect to light cone `time' $X^-$ has an RMS size in the conjugate light cone direction $X^+$ that is formally linearly divergent in mode number \cite{lennyspreading, BHpaper}.  This leads to an estimate for its size as measured by a detector $D$  which depends on its kinematics as
\begin{equation}
\Delta X^+_* \sim \alpha' p^+_D
\end{equation}
This light cone gauge estimate was supported in \cite{sixpoints, dilatontracer} by an on-shell six point function amplitude containing an off-shell
%\footnote{The necessity for an off-shell detector arises because we want it to have both a well-defined $p^+_D$ and to be localized in $X^+$. Formally, this is achieved by creating $D$ as an intermediate-string in a 6-point amplitude.} 
open string process $C+D \rightarrow \hat{C}+\hat{D}$. At a longitudinal separation $X^+_D - X^+_C \sim \alpha' p^+_D$,  this amplitude is parametrically stronger than the corresponding quantum field theory amplitude.

We will derive the analogous closed string estimate in Section \ref{TreeLevel}, which have an interaction strength growing with the Mandelstam invariant $s$ as $G_N s_{CD}\sim G_N p^+_D p^-_C$.   This can naturally be applied as in \cite{Backdraft}\cite{BHpaper}\cite{Danjie}\ to interactions that take place in the near-horizon Rindler region $X^+ X^- \ll r_s^2$ of a black hole with Schwarzschild radius $r_s$.  Here $C$ is a string dropped at a Schwarzschild time $t_0$, while $D$ is a detector lowered to the near-horizon region at some later time $t_0 + \Delta t$.  The evolution in the black hole background generates a large center of mass energy via the exponential growth of $p^+_D \propto e^{\Delta t/2r_s}$.  
Since $\Delta X^+_* \sim X^+_D - X^+_C$ grows at a similar rate, once $\Delta t$ exceeds $r_s$, the relation $\Delta X^+_* \sim X^+_D - X^+_C$ is equally easy to satisfy at larger $\Delta t$, implying that the detector can interact with $C$ at arbitrarily late times. However, as $p^+_D$ increases, so will the effective coupling constant $G_N s_{CD} \sim G_N p^-_C p^+_D$. Once we get into the regime $G_N s \gtrsim 1$, loop effects will become important and they will need to be taken into account.

While in certain regimes loop effects were found to suppress nonlocality \cite{grossmende}, in Section \ref{LoopCorrections} we will find that in the eikonal regime spreading persists unimpeded throughout the regime of perturbative control. By pushing $G_N s_{CD} \gg 1$ towards the limit of validity of the eikonal approximation, we obtain strong, semi-classical (but stringy) interactions that are dominated by a saddle point with momentum transfer

\begin{equation}
\Delta q_{\perp} \sim \frac{G_N s_{CD}}{x_{\perp}}\gg \frac{1}{r_s}
\end{equation}
where $x_{\perp} < r_s$ is the transverse separation. These deflections are sensitive to the exact kinematics of $C$ and $D$, and it becomes apparent that if we had perfect knowledge of the initial state of $D$, we could figure out something about the kinematics of $C$ by looking at the scattered state $\hat{D}$.

In Sections \ref{DetectorKinematics}\ and \ref{InfoRecovery}, we will extrapolate these $2 \rightarrow 2$ interactions to more general $N_C + N_D \rightarrow N_C + N_D$ interactions in a Schwarzschild black hole background.  We will show that $D$ can mine a large amount of information about the initial (pre-scattering) kinematics of the infalling strings $C$, although it is strongly sensitive to only half of the kinematic variables. To quantify this statement, we will introduce reference systems $R, A$ entangled with $C,D$ respectively, similarly to the Hayden-Preskill setup \cite{HaydenPreskill}.  We obtain from the effects that are simplest to detect, in the limit $S_D \ll S_C$ 
%(up to some logarithmic factors we will compute later)
\begin{equation}
I(\hat{D}A: R) \simeq S_D
\end{equation}
Conversely, in the opposite limit $S_C \ll S_D$ we obtain $I(\hat{D}A:R) \simeq S_C$ from the most easily detected effects.  However, in this regime with many detector states, more information may be detectable via subleading dependences on kinematic variables in the amplitude.  In this paper, we make no attempt to derive the optimal detector state and data analysis prescription, but the results 
%mirror Hayden-Preskill, and they 
we find indicate that substantial information about the kinematics of $C$ can be recovered via longitudinal spreading. 
%In Section \ref{InfoRecovery}, we will discuss the relation of our results to Hayden-Preskill, and the drama that the infaller experiences.

\subsection{Broader context:  why string theory?}

Before getting to our analysis, let us put this in a broader context.  The effect we derive, extending \cite{lennyspreading, BHpaper, sixpoints, dilatontracer, Backdraft, Danjie}\ manifests a striking behavior of perturbative string theory which goes beyond perturbative effective field theory.  We note that this may have other applications, e.g. to gauge theory scattering; the present analysis readily translates to holographic Yang Mills theory \cite{Warpedpaper}.    

The application we explore in this work is the black hole information problem.  For a review detailing various approaches see \cite{JoeReview}.  More recent progress includes a remarkable derivation of the Page curve in the bulk for solvable models \cite{PageCurveW}\cite{PageCurveE}, building from AdS/CFT developments in entanglement wedge reconstruction  which imply the Page curve via the duality \cite{EWreconstruction}\cite{Islands}.  This provides a  detailed test of unitarity, which is a general implication of AdS/CFT, while raising numerous additional technical and conceptual questions such as those studied very recently in \cite{MarolfMaxfield}.  

Still, it is necessary to analyze the real-time evolution of the system and identify the mechanism for information transfer from the matter that formed the black hole to the radiation that is left after it decays.  A complete calculation would determine the status of the putative near-horizon region during this process; for recent comments on the nontriviality of this see \cite{InfoCommentary}. A complete treatment would also determine the physics of the putative black hole singularity.
That is a daunting task for quantum gravity in, say, four large dimensions with a realistic value of the cosmological constant; it requires control of finite-entropy subsystems related to finite patches of spacetime at the appropriate level of approximation.  The holographic treatment of this, including entanglement wedge reconstruction, is under active development (as in e.g. \cite{TTEE}\ and references therein, see also \cite{PageCurveW}).  

In general, it is important to identify the leading contributions to bulk dynamics and capture their effects. 
The very recent developments just cited lend support to the idea that even exponentially small (gravitational instanton) effects $\sim e^{-r_s^2/G_N}$, in combination with the large number $e^{r_s^2/G_N}$ number of states, may be enough to account for the information transfer. 
Nonetheless, it remains crucial to identify the leading contribution to the dynamics in string theory, the leading candidate for a theory of quantum gravity in higher dimensions.  String theory contains additional scales, and generically corrects effective field theory before the Planck scale.   As such, in many contexts it determines the leading dynamics regardless of whether Planckian physics would in principle be enough.  

A related, general motivation is that quantum gravity in its known forms contains extended objects; there are some interesting hints that this kind of non-locality is inevitable, see e.g. \cite{JuanST}.  In string theory, this is essential to its mechanism for UV finite perturbative amplitudes, a necessary feature of quantum gravity.  In some cases relevant to the approach we pursue in this work, the string theoretic amplitude is not obtained from a convergent expansion in the corresponding quantum field theory amplitude \cite{sixpoints}:  it is not sufficient to treat the dynamics in a perturbative $\alpha'$ expansion.  

The need for string-theoretic or quantum gravitational effects is clearest in the problem of spacetime singularity resolution.  Numerous examples are known where perturbative string theory resolves the singularity before the system enters into a Planckian regime \cite{STtimelike}, and others where quantum physics is required \cite{Plancktimelike}.  Examples where string theory intercedes include certain spacelike singularities including BTZ black hole singularities \cite{STspacelike}.  The set of string theoretic examples includes topology changing processes including `baby universe' formation \cite{STtimelike}, in a regime where the analogous Euclidean quantum gravity effect is not applicable.  For example, in a system with a Scherk-Schwarz circle with antiperiodic Fermion boundary conditions, at small radius a winding string `tachyon' goes unstable and condenses, whereas at large circle radius the exponentially suppressed Euclidean `bubble of nothing' instanton \cite{Wittenbubble}\ pertains.  This may be viewed either as a bug or a feature:  it indicates that the physics is non-universal, but on the other hand the technical analysis of perturbative string theory effects is simpler than full-fledged quantum gravity.\footnote{It may facilitate an understanding of the latter as well via correspondence limits analogous to \cite{correspondence}.}       

For horizon physics, similarly to the physics of singularities, there are regimes  where perturbative string theory effects take the system beyond the effective field theory prediction before the Planckian regime, as a result of the relative boost reviewed above that is generated near black hole horizons \cite{Backdraft}\cite{BHpaper}.  It is that regime which we analyze in the present work.  Moreover, our methods apply to the Lorentzian target spacetime signature and are simplest in asymptotically Minkowski spacetime.  

\subsection{The AdS/CFT case}

Before coming to our detailed analysis, we note a potential subtlety with the effect we obtain in this work.\footnote{We thank D. Stanford for discussions of this.}  One might be tempted to apply the same result to a Rindler horizon in AdS spacetime.  In that context, the late time behavior of correlators of local operators is bounded, which seems in tension with the interactions we find between early infallers and late detectors near a black hole horizon. However there is not a sharp conflict between that and the present analysis, for a number of reasons including the fact that there is a time limit on the validity of the Eikonal approximation in our setup;  it is also somewhat nontrivial to relate near-horizon string states to boundary operators (see also the discussion in the final section of \cite{sixpoints}).  
The progress of thermalization at finite temperature is also impacted by effects like ours, but again it is difficult to compare to the CFT because there is not a controlled analysis of this on the field theory side of large-radius AdS/CFT.  
In any case, it would be interesting to carry out a detailed analysis of this effect in AdS spacetime, and translate it into the appropriate dual CFT observables.  

In this regard, we should reiterate that in the present work we do not analyze interactions of $D$ with near-horizon Unruh radiation. If that were a large enough effect to destroy $D$'s transport of kinematical information of the on-shell $C$ infaller, it would be an interesting (and rather surprising) string-theoretic effect in itself.  Again, we do not know of a controlled calculation from AdS/CFT that would contradict our finite-time information recovery effect via the string spreading induced $C$-$D$ interaction.

\section{Closed String Spreading at Tree-Level}\label{TreeLevel}

In this section, we will demonstrate that the closed string S-matrix exhibits longitudinal spreading at tree-level, generalizing the results of \cite{sixpoints, dilatontracer}. While transverse spreading can be obtained at the level of 4-point functions, longitudinal spreading is more elusive. An on-shell $2\rightarrow 2$ process is fully parametrized by the center-of-mass energy $s \sim E_{\text{c.o.m.}}^2$, and the impact parameter $x_{\perp}$ (or equivalently the transverse momentum transfer $t \sim \Delta q_{\perp}^2$). To understand this picture, imagine a string $C$ moving along the $X^-$ direction which is localized around $X^+_C$, and a string $D$ moving along the $X^+$ direction which is localized around $X^-_D$. 

\begin{figure}[!tbp]
  \centering
  \subfloat[A $2\rightarrow 2$ scattering event.]{\includegraphics[width=0.5\textwidth]{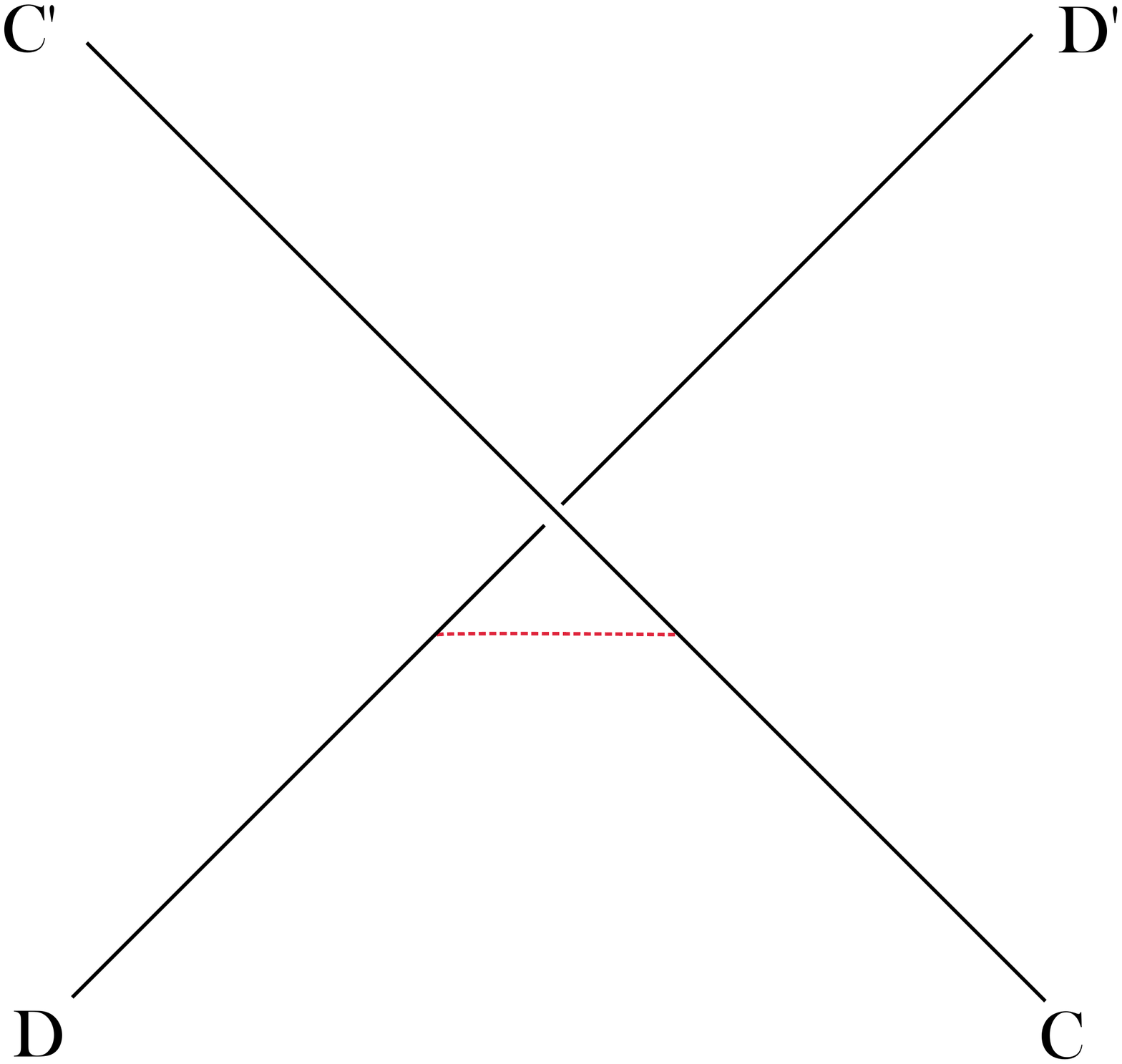}\label{fig:Transverse}}
  \hfill
  \subfloat[A $3\rightarrow 3$ scattering event.]{\includegraphics[width=0.5\textwidth]{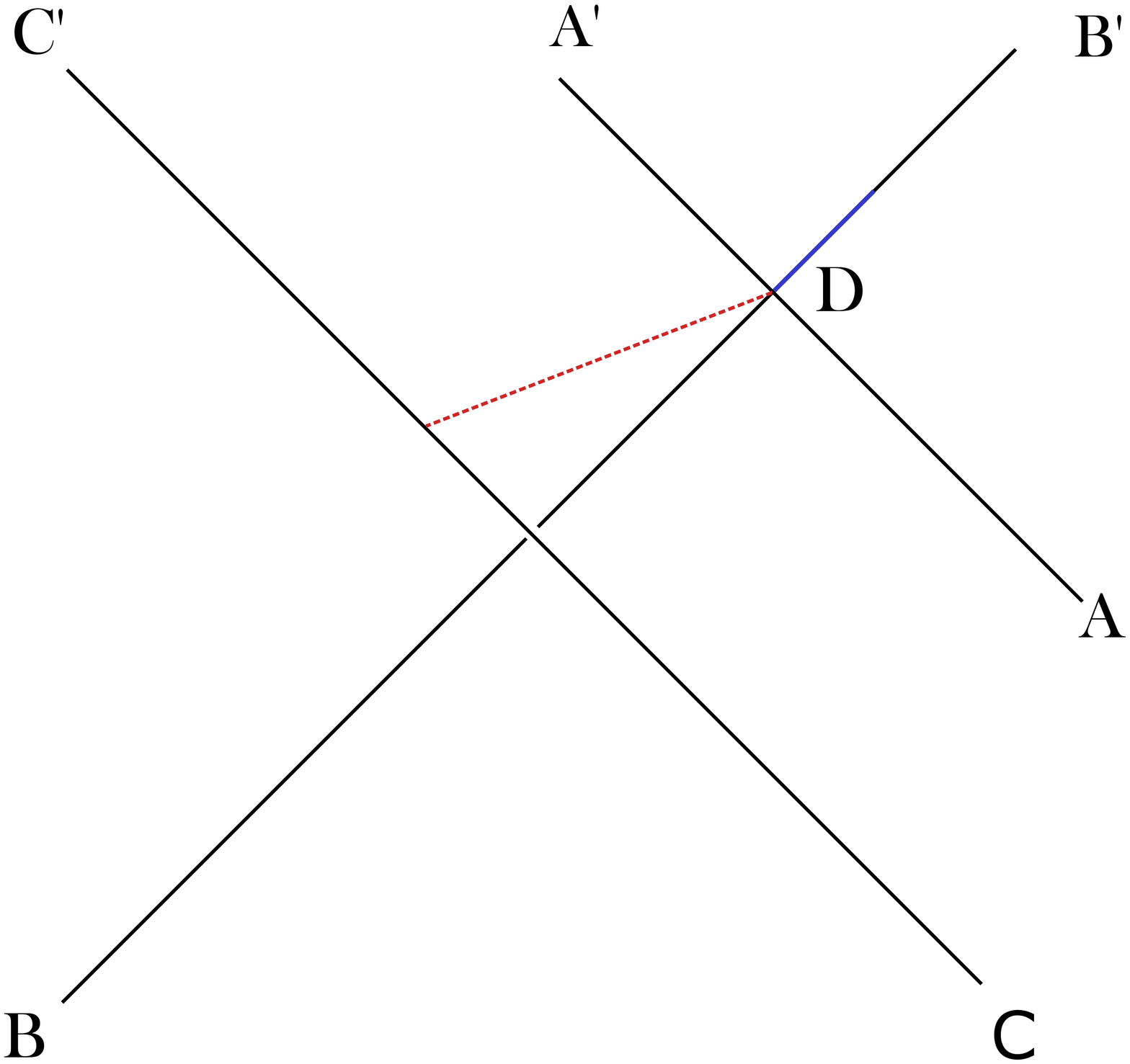}\label{fig:Longitudinal}}
  \caption{In a purely $2\rightarrow 2$ scattering event, it is
  difficult to test longitudinal interactions. By using an auxiliary string $A$, we can create a detector $D$ that is longitudinally localized \cite{sixpoints}.}
\end{figure}

The longitudinal coordinates of these strings will meet at $(X^+_C, X^-_D)$, and the only possible separation left is transverse (see Figure \ref{fig:Transverse}). In order to achieve longitudinal separation, we would need $D$ to spontaneously appear at some finite $X^+_D > X^+_C$. To achieve this, we will treat $D$ as an off-shell, intermediate string that appears in a 6-point amplitude $A+B+C \rightarrow \hat{A}+\hat{B}+\hat{C}$ (see Figure \ref{fig:Longitudinal}). In an appropriate kinematic regime the 6-point amplitude will factorize into sub-processes $A+B \rightarrow \hat{A}+D$ and $C+D \rightarrow \hat{C}+\hat{B}$. The auxilliary string $A$ is chosen to be localized at some $X^+_A > X^+_C$ (and moving in the $X^-$ direction), while $B$ is localized at $X^-_B$ and moving in the $X^+$ direction\footnote{In the near-horizon setup we consider, the process of lowering a detector at a time $t_D \gg t_C$ plays the role of a larger amplitude that creates a detector at some finite $X^+_D$.}.

When $A$ and $B$ interact, the intermediate string $D$ will be localized at $(X^+_A, X^-_B)$ and its interaction with $C$ will have a well-defined longitudinal separation. To obtain evidence for spreading, we will follow \cite{sixpoints} and show that the 6-point amplitude $\mathcal{A}_{A+B+C \rightarrow \hat{A} + \hat{B} + \hat{C}}$ has significant support at $X^+_A \gg X^+_C$. By this we mean that the amplitude does not arize from scattering off the tails of localized wave-packets. If $X^+_A$ and $X^+_C$ are Gaussianly distributed, then local scattering off the tails of their wavepackets will give a suppression factor $\sim e^{-(X^+_{A,0} - X^+_{C,0})^2/\sigma^2}$. We seek to find stringy contributions that aren't suppressed in this regard.

This property is defined in position space (with a use of appropriately localized wavepackets for our strings), but it also manifests in the phase of the momentum-space amplitude. Recall that multiplying a momentum-space wavefunction $\psi(p)$ with a phase-factor $e^{-ipa}$ leads to a shift $x \rightarrow x+a$ (up to factors of $2\pi$) in the position-space wavefunction. In our problem, the corresponding statement is that if the momentum-space amplitude $\mathcal{A}(K_{IJ})$ (where $K_{IJ} = 2\alpha' k_I \cdot k_J$ denote the Mandelstam invariants) can be written as
\begin{equation}
\mathcal{A}(K_{IJ}) \sim e^{-i k_C^- \Delta X^+_*} \mathcal{A}_{slow}
\end{equation}
where $\mathcal{A}_{slow}$ has a slow dependence on $k_C^-$. The above factor would effectively shift the position-space wave-packet of $C$ by $\Delta X^+_*$, and for $X^+_A - X^+_C \simeq X^+_D - X^+_C \simeq \Delta X^+_*$ we would be scattering off the center of the Gaussian wave-packets, not the tails. Operationally, we can test the above prediction by considering narrow wave-packets for $A,C$ with uncertainties $\Delta X^+_A, \Delta X^+_C \ll \Delta X^+_*$ and then investigate whether the amplitude $\mathcal{A}(X)$ exhibits parametric enhancement compared to appropriate field theoretic comparison models with the same wavepacket. 

In the next subsection, we will find that the 6-point amplitude in position space is given (in an appropriate Reggeized regime) by the saddle point

\begin{equation}
\mathcal{A}_6 \sim \mathcal{A}_{A+B \rightarrow \hat{A} + D} \times \frac{g_s^2 (\alpha' s_{CD})^{2+\alpha' t_{C\hat{C}}/4}}{\alpha' t_{C\hat{C}}}\times e^{i\pi \alpha' k_{C\hat{C}}^- k_{\hat{B}}^+} e^{i\pi \eta} \frac{2^{\eta} }{\sin \pi \frac{\eta}{2}} e^{\frac{1}{2|\eta|} (\alpha' k_D^2 + \eta+4)^2}
\end{equation}
In the above, $s_{CD}$ is the center-of-mass energy of the $C+D \rightarrow \hat{C}+\hat{B}$ sub-process and $t_{C\hat{C}} \ll s_{CD}$ is the momentum transfer. The variable $\eta$ is a function of the transverse momenta, which we take to be large and negative with $\eta \simeq -\alpha' k_D^2$ so that the above saddle point will be valid. 
%{\color{blue} Add a brief discussion of Gaussian wavepackets}
Details on the above amplitude and its derivation are given in Section \ref{TreeDetails}, but the reader can skip ahead to Section \ref{LoopCorrections}.  As in \cite{sixpoints}, we employ Gaussian wave packets.  

Once we have obtained the amplitude, the probability for interaction is given by
\be\label{Prob}
Prob \sim\sum_{final~states}\frac{1}{N_{id}!} \int \prod_f \frac{dp_f d^{d-2}q_f}{2 E_f} |{\cal A}(X)|^2
\ee
where $N_{id}$ is the number of identical outgoers. After keeping track of all the kinematic factors coming from the amplitude and the wave-packets, we will see that the probability of interaction grows with the relative longitudinal boost of $C$ and $A,B$. At tree-level, the probability will scale as
\begin{equation}
\text{Prob}_{tree} \sim s^2_{CD} G_N^2
\end{equation}

In the black hole context, this shows that the probability of interaction will grow exponentially with the relative boost $e^{\Delta t/2r_s}$ until we reach the regime $G_N s_{CD} \sim 1$. As the probability of interaction grows to $O(1)$, loop effects must be taken into account to prevent the probability from exceeding 1.

\subsection{Tree-Level Amplitude}\label{TreeDetails}

We will start with the elastic momentum-space six point amplitude at tree level.  The origin of the spreading interaction within the open string version of this amplitude was explained relatively simply at the level of vertex operator integrals in \S 4.3\ of \cite{sixpoints}.  Compared to that, we would like to make two generalizations.  

First, we will work with closed strings instead of open strings because closed string interactions dominate in our amplitude as we increase $G_N s$.  Secondly, we will work with $K_{C\hat C}\ll 1$, near a graviton pole, rather than $K_{A\hat A}\ll 1$. It is in this regime that the amplitude and probability grow with time (as we will see shortly), and it is most suited for generalization to the eikonal amplitude. For simplicity, we take the external legs to be bosonic string tachyons, since the growing interaction is mediated by internal legs including the gravitons at the massless level.  

Let us use $SL(2,C)$ to set $z_{\hat B}=0, z_B=\infty, z_{A}=1$.  The amplitude is
\be\label{Vcorr}
\mathcal{A}_6 \sim g_s^4 \langle \int d^2 w V_C (w)\int d^2z_{C\hat C} V_{\hat C}(z_{C\hat C}) V_A(1)\int d^2 z_{\hat A} V_{\hat A} ~ V_B(\infty) V_{\hat B}(0) ~ \rangle \langle \prod c\bar c\rangle
\ee
in terms of the integrated vertex operators, where the last factor is the $c$ ghost correlator \cite{JoeBook}.  
We will make use of the method introduced by Brower et al \cite{BPST}, replacing $V_C$ and $V_{\hat C}$ by the Pomeron vertex operator that appears in their generalized OPE:  
\be\label{PomV}
\int d^2 w V_C V_{\hat C} \simeq \frac{2\pi \Gamma(-1 -\alpha' t_{C\hat C}/4)}{\Gamma(2+\alpha' t_{C\hat C}/4)}e^{-i\pi(1-\alpha' t_{C\hat C}/4)}e^{i k_{C\hat C}X(z_{C\hat C})}\left[k_C\cdot\partial X ~ k_{\hat C}\cdot\bar\partial X(z_{C\hat C}) \right]^{1+\alpha' t_{C\hat C}/4}
\ee  
where $k_{C\hat C}=k_C+k_{\hat C}$.  This is
valid in the Regge regime, $\alpha' t_{C\hat C}\ll K_{C\hat B}=\alpha' s_{CD}$, and we will take the intermediate string $C\hat{C}$ to be near the graviton pole $t_{C\hat C} \simeq 0$. The amplitude (\ref{Vcorr}) gets its leading contribution from contractions of the Pomeron with the $\hat B$ vertex operator, yielding
\be\label{FirststepA}
\sim g^2_s \frac{(\alpha' s_{CD})^{2+\alpha' t_{C\hat C}/4}}{\alpha' t_{C\hat{C}}}\int  d^2 z_{C\hat C} d^2 z_{\hat A}
 |z_{C\hat{C}}|^{\alpha' k_D^2/2- \alpha' t_{C\hat C}/4}\, |z_{\hat{A}}|^{K_{\hat B \hat A}/2 }|1-z_{C\hat{C}}|^{K_{C\hat{C} A}/2}|1-z_{\hat A}|^{K_{A\hat{A}}/2}|z_{\hat A}-z_{C\hat{C}}|^{K_{C\hat{C}\hat A}/2} 
 %\times K_{polarization}(\{z\}, \{k\})
\ee 
where in the first factor in the integrand we noted that $D=C +\hat C+ \hat B$ (with the other terms in the exponent coming from the contraction of $(\partial X\bar\partial X)^{1-\alpha' t_{C\hat C}/4}$ with $e^{i k_{\hat B}X}$). We can perform a similar Pomeron analysis for the $A\hat A$ string. As we have chosen $A$'s kinematics so its momentum is primarily in the $p^-$ direction, the main contribution to the amplitude will come from contractions of the $A\hat A$ Pomeron with the $\hat B$ vertex operator, which will yield
\be\label{nextstepA}
\sim \frac{(\alpha' s_{CD})^{2+\alpha' t_{C\hat C}/4}}{\alpha' t_{C\hat C}}\frac{\Gamma(-1-\alpha' t_{A\hat A}/4)}{\Gamma(2+ \alpha' t_{A\hat A}/4)}(\alpha' s_{AB})^{2-\alpha' t_{A\hat A}/4} \int d^2z_{C\hat C}  |z_{C\hat{C}}|^{\alpha' k_D^2/2}\,|1-z_{C\hat{C}}|^{\alpha' k_{C\hat{C}}\cdot( k_A+k_{\hat A})}
\ee   
where we dropped the $\alpha' t_{C\hat C}$ in the exponent of $|z_{C\hat C}|$ since we are near the graviton pole. Defining 
\be\label{etadef}
\eta \equiv \alpha' k_{C\hat{C}}\cdot( k_A+k_{\hat A})
\ee
we see that the amplitude has factorized into a product of four point amplitudes times a factor taking the same form as a Virasoro-Shapiro amplitude, with arguments $\simeq k_D^2\alpha'/2$ and $\eta$.  
\bea\label{Fdef}
F(\alpha' k_D^2/2, \eta) &=& \frac{2\pi \Gamma(\alpha' k_D^2/4+1)\Gamma(1+\eta/2)\Gamma(-1-\alpha' k_D^2/4-\eta/2)}{\Gamma(\alpha' k_D^2/4+\eta/2+2)\Gamma(-\eta/2)\Gamma(-\alpha' k_D^2/4)}\nonumber\\
&=& 2\frac{\sin\pi \frac{\alpha' k_D^2}{4}\sin\pi(\frac{\alpha' k_D^2}{4}+\frac{\eta}{2})}{\sin\pi \frac{\eta}{2} }\frac{\Gamma(1+\frac{\alpha' k_D^2}{4})^2\Gamma(-1-\frac{\alpha' k_D^2}{4}-\frac{\eta}{2})^2 }{\Gamma(\frac{-\eta}{2})^2} \nonumber\\
\eea
Similarly to the open-string amplitude analyzed in \cite{sixpoints}, the phase structure of this factor leads to spreading in a kinematic window $0< \alpha' k_D^2< -2\eta$, in which the second form of (\ref{Fdef}) is useful, with all arguments of the $\Gamma$ functions positive.  This expression has a term with a phase
\be\label{phasefactor}
e^{i\pi \alpha' k_D^2/2} \simeq e^{i \pi k_{\hat B}\cdot k_{C\hat C}+\dots}
\ee  
multiplying a factor that has an extremum at $\alpha' k_D^2=-\eta-4$.   With a thin wavepacket for $k_C^-$ in momentum space, this strongly varying phase leads to a contribution that is peaked in position space at the spreading radius

\begin{equation}\label{Xplusstar}
X^+_* = 2 \pi \alpha' E_{\hat{B}} = \sqrt{2} \pi \alpha' k_{B}^+
\end{equation}
More specifically, at $\alpha' k_D^2 \simeq -\eta-4$ we get

\begin{equation}
F(\alpha' k_D^2/2, {\eta}) \sim e^{i\pi \alpha' k_{C\hat{C}}^- k_{\hat{B}}^+} e^{i\pi {\eta}} \frac{2^{{\eta}} }{\sin \pi \frac{{\eta}}{2}} e^{\frac{2}{|\eta|} (\frac{\alpha' k_D^2}{2} + \frac{\eta}{2}+2)^2}
\end{equation}
Putting everything together we have
\begin{equation}\label{Momentum6point}
\mathcal{A}_6 \sim \mathcal{A}_{A+B \rightarrow \hat{A} + D} \times \frac{g_s^2 (\alpha' s_{CD})^{2+\alpha' t_{C\hat{C}}/4}}{\alpha' t_{C\hat{C}}}\times e^{i\pi \alpha' k_{C\hat{C}}^- k_{\hat{B}}^+} e^{i\pi \eta} \frac{2^{\eta} }{\sin \pi \frac{\eta}{2}} e^{\frac{1}{2|\eta|} (\alpha' k_D^2 + \eta+4)^2}
\end{equation}

Before we proceed, we want to note that the $2^{\eta} \ll 1$ factor has a natural interpretation in terms of the light-cone prediction, which was

\begin{equation}
\Delta X^+_{spr} \sim \frac{k^+_D}{k^2_D}
\end{equation}
As reviewed in \cite{sixpoints}, the light-cone result fits with \ref{Xplusstar} if we assume that at $\Delta X^+ > \Delta X^+_{spr}$ the string follows an exponential distribution

\begin{equation}
e^{-c \Delta X^+/\Delta X^+_{spr}}
\end{equation}
for some numerical constant $c$. Then, since $X^+_*/\Delta X^+_{spr} \sim \alpha' k_D^2 \sim -\eta$, an exponential suppression in $\eta$ is expected, and it indeed manifests in the $2^{\eta}$ factor.

In Section \ref{LoopCorrections}, we will analyze loop corrections to this result, focusing on the fate of the phase factor (\ref{Fdef}) that leads to spreading. The perturbative string analysis will break down by the time the non-local kinematic variable $s_{CD}$ has grown so much that the detector and the part of $C$ with which it interacts are within their own Schwarzschild radius.  

%$\bullet$ with kinematic constraints as in our recent notes, anticipating relevant regime for the black hole context (fixing early kinematics etc), full probability at tree level.

\subsection{Comparison with Field Theory}\label{Comparison}

In momentum space, our amplitude has a phase factor which demonstrates the possibility for string spreading. However, this phase factor is only clear near the extremum $\alpha' k_D^2 \simeq -\eta$, and thus it is necessary to use wavepackets that localize this interaction near this extremum, while also demonstrating a clear longitudinal separation. These requirements of localization in both position and momentum space are conflicting, but as in \cite{sixpoints} we will find a regime where both conditions can be satisfied and where the string theoretic amplitude is parametrically stronger than an EFT estimate.

Following \cite{sixpoints}, we will convolve our momentum-space amplitude with Gaussian wavepackets $\Psi_A^+(X^+_A), \Psi_C^+(X^+_C)$ whose widths $1/\sigma_A, 1/\sigma_C$ (we will use $\sigma$ to refer to momentum-space widths) are much smaller than the central separation $X^+_{A_0} - X^+_{C_0}$. In the limit of large boosts, the width $\sigma_A \sim e^{\kappa}$ will be much larger than $\sigma_C$, and also $X^+_{A_0} \gg X^+_{C_0}$ so we will require $1/\sigma_A \ll X^+_{A_0} \simeq 2  \pi \alpha' E_{\hat{B}}$.

A second requirement is that we want our kinematics to remain near the extremum $\alpha' k_D^2 \simeq -\eta-4$. In momentum space, we saw that our amplitude \ref{Momentum6point} grows as $e^{(\alpha' k_D^2 + \eta + 4)^2/2|\eta|}$, so we need a sufficiently strong Gaussian suppression to ensure that the dominant contribution comes from near the extremum. Following Section 5 of \cite{sixpoints}, we can write the momentum-space width of $A$ as

\begin{equation}
\sigma_A = \frac{\sqrt{c_\sigma (-\eta)}}{4 E_{\hat{B}} \alpha'}, \quad c_\sigma < 1/2
\end{equation}

Besides ensuring that the extremum becomes a local maximum after we include the wavepacket suppression, we make the stronger demand that it is a global maximum, i.e. that the amplitude is suppressed near the poles $k_D^2 = 0$ and $k_D^2 = -\eta-4$. This will ensure that we do not have to worry about contributions from the poles. The wavepacket suppression is $e^{\eta/2c_\sigma}$, and thus we need

\begin{equation}
c_\sigma < \frac{1}{2 \log(2)}
\end{equation}
The last demand we make is that our spreading-induced amplitude is larger than a field-theoretic amplitude (with the auxiliary process $A+B \rightarrow \hat{A}+D$ stripped off)
\begin{equation}
\mathcal{A}_{QFT} \sim e^{-X^2 \sigma_A ^2/2} \frac{\lambda_{CD \hat{C}\hat{B}}}{-\eta/2}
\end{equation}
If we identify $\lambda_{CD \hat{C}\hat{B}} = \mathcal{A}_{B+C \rightarrow \hat{B}+\hat{C}}$, then as $-\eta \gg 1$ we have 
\begin{equation}
\frac{\mathcal{A}_{QFT}}{\mathcal{A}_{ST}} \simeq \frac{1}{-\eta/2}e^{-X^2 \sigma_0^2/2 - \eta \log(2)} = \frac{1}{-\eta/2}e^{\eta(\frac{\pi^2 c_\sigma}{8}-\log(2))}
\end{equation}
This imposes the constraint $c_\sigma > 8\log(2)/\pi^2$, and thus we have the window
\begin{equation}
\frac{8\log(2)}{\pi^2} < c_\sigma < \frac{1}{2\log(2)}
\end{equation}
This window is entirely analogous to the open string case, except it is rescaled by a factor of 4 (which arises from the fact that $X^+_{*, closed} = 2\pi \alpha' E_{\hat{B}}$ while $X^+_{*, open} = 4\pi \alpha' E_{\hat{B}}$).

\subsection{Tree-level probability}\label{sec:probability}

At this point, it is worth estimating the full probability for scattering at tree level.  This brings in the final state phase space integrals and integrals over wavepackets that go into (\ref{Prob}).  We must only keep contributions which maintain the kinematics leading to the factor of (\ref{Fdef}).    

Let us first consider the final state phase space.  We have
\be\label{fphasesp}
\frac{\Delta p_{\hat A} \Delta p_{\hat B}\Delta p_{\hat C}  (\Delta q_{\hat A})^2 (\Delta q_{\hat B})^2 (\Delta q_{\hat C})^2}{8 E_{\hat A} E_{\hat B} E_{\hat C}}
\ee
One important limiting factor is the following.  The spreading effect is manifest in our S-matrix amplitude for a limited range of $k_D^2=4 E_{\hat B}(E_{\hat C}-E_C)+\dots$ around $-\tilde \eta/2$  (see \cite{sixpoints} for a detailed account of the kinematics).  Although this is somewhat larger than 1 in the controlled regime described in \cite{sixpoints}, the effect becomes suppressed exponentially in $\tilde \eta$, so we must keep $k_D^2\alpha'$ from growing too large.  Since $\hat B$ is part of the late system, its energy gets boosted up: 
\be\label{kapdef}
E_{\hat B}= E_0 e^{\Delta t/2 r_s}\equiv E_0 e^\kappa.
\ee
As a result, we must limit 
\be\label{phatcrange}
\Delta p_{\hat C} \sim \frac{k_{D}^2}{E_0}e^{-\kappa}   
\ee
The other final state phase space factors are not parametric in $e^\kappa$.  To remain in the spreading window of kinematics, with $K_{C\hat C}\sim \delta q^2 \ll 1$,  we require $\Delta q_{\hat C}\sim \delta q, \Delta q_{\hat A}\sim \Delta q_{\hat B} \ll q$.  Altogether, we find that the final phase space factors (\ref{fphasesp}) scales like
\be\label{fphasefinal}
\prod_f \frac{d^3\p_f}{E_f}\sim \frac{e^{-\kappa} k_D^2}{E_0 E_{\hat C}} \frac{\Delta p_{\hat B}}{E_{\hat B}}\frac{\Delta p_{\hat A}}{E_{\hat A}} q^4 \delta q^2
\ee        
Noting that $\tilde\eta \sim q\delta q\alpha'$, the last two factors here scale like $q^2\tilde{\eta}^2/\alpha'^2$.  

We have
\be\label{Apsi}
{\cal A}(X^+) = \int \frac{\prod_I {dp_I d\q_I}}{\sqrt{2 E_I}}\Psi_I(p_I, \q_I) \delta(\sum p_{\hat I}-\sum p_I)\delta^{\perp}(\sum \q_{\hat I}-\sum q_I)\delta(\sum\omega_{\hat I}-\sum\omega_I) ~  \hat{{\cal A}}
\ee
with wavepackets that depend on their peak positions, including $X^+$, the separation between incomers A and C.  These wavepackets take the form 
\be\label{Psiform}
\Psi_I \sim \frac{1}{\sigma_{Ip}^{1/2}\sigma_{Iq}} e^{-\delta p_I^2/2\sigma_{Ip}^2}e^{-\delta q_I^2/2\sigma_{Iq}^2} \times phase
\ee
where $\delta q_I, \delta p_I$ here denotes the variation of the momentum from the value it takes at the peak of the wavepacket.  

Let us use the energy delta function to do the $p_B$ integral, and the spatial momentum delta function to do the $p_A, q_A$ integrals.  In particular, in the longitudinal direction we have $\delta p_A\sim -\delta p_C$.  In the transverse direction we take $\sigma_{Cq}\sim \delta q$ and $\sigma_{Aq}\sim \sigma_{Bq} \ll q$.  The amplitude becomes
\be\label{Apsinext}
{\cal A}(X^+) \sim  \frac{ {\delta q^2 \sigma_{Bq}^2}}{\sqrt{8 E_{A0}E_{B0}E_{C0}}}\frac{1}{\sqrt{\sigma_{Aq}^2\sigma_{Bp}\sigma_{Bq}^2\sigma_{Cq}^2}} \int \frac{d\tilde p_C}{\sqrt{\sigma_{Ap}\sigma_{Cp}}}e^{-\delta\tilde p_C^2(\frac{1}{\sigma_{Cp}^2}+\frac{1}{\sigma_{Ap}^2})}e^{i \tilde p_C^- X^+}  ~  \hat{{\cal A}}
\ee
Let us for simplicity take all the local energy scales of the same order, $\sim E_0$, and take $\sigma_{Ap}=\sigma_{Cp}$. 

In (\ref{Apsinext}), $\hat{\cal A}$ is the momentum-space amplitude.   At tree level, as we have described above, this is
\be\label{Ahatrreefull} 
{\hat {\cal A}}_{tree}=\frac{K_{C\hat B}^{2-K_{C\hat C}/2}}{K_{C\hat C}}K_{\hat B\hat A}^{-K_{A\hat A}/2} F(k_D^2\alpha', \tilde\eta)
\ee
(We will also estimate the probability in the generalization to an analogue of the Eikonal regime below.)
The amplitude then scales like
\be\label{Apsisize}
{\cal A}(X^+) \sim E_0^{-3/2} \delta q \frac{\sigma_{Bq}}{\sigma_{Aq}\sqrt{\sigma_{Bq}}}\hat{\cal A}
\ee
where $\hat {\cal A}$ is evaluated at the peak $p_C$ momentum in the spreading regime.   At tree level, this scales like 
\be\label{Atreescale}
\hat{\cal A}_{tree}\sim \frac{E_0^4 e^{2\kappa}G_N}{K_{C\hat C}} 2^{\tilde\eta} \hat{\cal A}_{aux}
\ee

Putting together the final phase space factors (\ref{fphasefinal}) with $|\hat{\cal A}|^2$ yields a tree-level probability that scales like
\be\label{probtree}
Prob_{tree}\sim \frac{\sigma_{Bq}^2}{\sigma_{Aq}^2}  \frac{\Delta p_{\hat B}}{E_{\hat B}}\frac{\Delta p_{\hat A}}{E_{\hat A}} e^{3\kappa} E_0^4 G_N^2 \frac{E_0}{\sigma_{Bp}} \frac{q^4\alpha'}{E_0^2}2^{2\tilde\eta} \tilde\eta
\ee
where we used that in our regime $k_D^2\sim -\tilde\eta/2$ (and we drop order 1 factors here).  

We have not yet chosen the scale of $\sigma_{Bp}$.  If we embed our flat-space process in the near horizon region, 
\be\label{nearhregion}
\Delta X^+ \Delta X^- \ll r_s^2
\ee
this is constrained by requiring that $B$ be localized in that region during the entire process, which involves an $X^+$ range of order $E_{\hat B}\alpha' \sim E_0 e^\kappa \alpha'$      (\ref{Xplusstar}).   This requires a small width for B in the $X^-$ direction, and hence a large momentum-space width:  $\sigma_{Bp}\sim e^\kappa E_0\alpha'/r_s^2$.  Incorporating that yields
 \be\label{probtreenext}
Prob_{tree}\sim \frac{\sigma_{Bq}^2}{\sigma_{Aq}^2}  \frac{\Delta p_{\hat B}}{E_{\hat B}}\frac{\Delta p_{\hat A}}{E_{\hat A}} e^{2\kappa} E_0^4 G_N^2  \frac{q^4 r_s^2}{E_0^2}2^{2\tilde\eta} \tilde\eta \sim s_{early-late}^2 G_N ^2 \times f(q, E_0, r_s, \tilde\eta, \sigma_I, \dots) 
\ee
where $f$ does not depend on $\Delta t$.  

Thus we see that the probability for the closed string spreading-induced beyond-EFT interaction increases with $\Delta t$ at least until quantum corrections become important.

\section{Loop corrections to Spreading}\label{LoopCorrections}

Having established longitudinal spreading at the tree level, we now seek to incorporate higher loop corrections by resumming the eikonal series, following the methods of \cite{ACV, Giddings, MuzinichSoldate, Kabat}. In pure gravity, the eikonal series captures the leading contribution at large $s$ and fixed momentum transfer $t = -\Delta q_{\perp}^2$, and in it amounts to summing over ladder diagrams. Other contributions are subleading by factors of $t/s$, so they can be ignored when $t \ll s$. As the scattering angle in the center of mass frame is $\theta \sim \sqrt{-t/s}$, the eikonal series is valid for small angle scattering.

In impact parameter space, the eikonal series captures the behavior of gravity at large $s$ and impact parameter $x_{\perp} \gg r_s(s) = G_N \sqrt{s}$ (in 4 dimensions). Here, $r_s(s)$ is the Schwarzschild radius of a black hole with mass $\sqrt{s}$, so the eikonal series is valid as long as we stay away from kinematic regime where the interacting particles can form a black hole. In impact parameter space, the eikonal amplitude takes a simple form

\begin{equation}\label{EikonalEq}
\mathcal{A}_{\text{eikonal}} \sim s(e^{i\chi(s, x_{\perp})} - 1)
\end{equation}
where $\chi(s, x_\perp)$ is called the eikonal phase, and it equals $G_N s \log(\mu x_{\perp})$ in 4 dimensions ($\mu$ is an IR cutoff). The eikonal phase is responsible for two physical effects: a transverse momentum transfer and the Shapiro time-delay. The former is given by
\begin{equation}\label{EikonalMomentumTransfer}
\Delta q_{\perp} \sim \frac{\partial \chi}{\partial x_\perp}\sim \frac{G_N s}{x_\perp}
\end{equation}
while the latter is given by

\begin{equation}\label{EikonalTimeDelay}
\Delta X^-_D \sim \frac{\partial \chi}{\partial p^+_D} \sim G_N p^-_C \log(\mu x_\perp)
\end{equation}
These effects are semi-classical, and they can be reproduced by a shockwave calculation \cite{tHooft}. 

In string theoretic $2\rightarrow 2$ processes, \cite{ACV} found that the eikonal series still captures the high energy behavior just like in gravity. The only additional complication is the possibility of diffractive excitations, or inelastic string production. However, both of these effects happen at relatively small impact parameters $x_{D} = l_s \sqrt{G_N s}$ and $x_{I} = l_s \sqrt{\log(\alpha' s)}$ respectively. For larger $x_{\perp}$, the interaction is analogous to gravity.

In Section \ref{ResumEikonal} we will generalize the eikonal series to spreading-induced interactions, and we will find the exact same structure as in \ref{EikonalEq}. While the computation of the eikonal phase turns out to be difficult in general, we will be able to compute it in the regime where $k_{A\hat{A}}$ lies in a compactified direction with length $R$ (e.g. the internal sphere in AdS). In this scenario, the non-zero values of $\eta =\alpha' k_{A\hat{A}} \cdot k_{C\hat{C}} < 0$ arise from the exchange of Kaluza-Klein modes with mass $n/R$ (as opposed to the exchange of gravitons in the standard eikonal). We find that the lightest mode dominates, and the eikonal phase can be expressed as 
\begin{equation}
\chi(x_\perp, \frac{x^+}{\alpha' p^+_D}) \sim \frac{G_N s_{CD} e^{-x_\perp/R}}{x_\perp^{d-4}} \tilde{F}(\frac{x^+}{\alpha' p^+_D}, -n_A/R^2)
\end{equation}
where $\tilde{F}$ is a boost-independent function whose magnitude is $O(2^{\eta})$. From the above equation, we will derive formulas for the momentum transfer in the eikonal interaction that are entirely analogous to the gravitational case. In the same vein, just like in pure gravity, our eikonal series will be valid as long as we stay away from the "black hole formation" regime, i.e. we maintain $x_{\perp} \gg G_N \sqrt{s}$.

In the rest of this section, we will systematically derive the eikonal amplitude for longitudinal string spreading. In Section \ref{DetectorKinematics} we will relate our results on string spreading, which are expressed in terms of the 6-point amplitude, to the infaller-detector setup in the near-horizon region of a black hole, and we will examine its kinematic constraints.

\subsection{Formal Resummation}\label{ResumEikonal}

Let us next generalize to the Eikonal regime \cite{ACV, Giddings, MuzinichSoldate, Kabat}. 
%For this, we generalize the tree-level diagram.  
We fractionate the $C$ and $\hat C$ leg into $n$ Pomeron legs $(C\hat C)_i$, $i=1,\dots n$, each with a Pomeron vertex operator (\ref{PomV}).  These rungs of the ladder diagram will propagate between the ladder sides $C\to \hat C$ and $D\to \hat B$, summing over all possible crossings. We will then re-sum the series over all $n$ to obtain an exponential resummation.    

\begin{figure}
\begin{center}
\includegraphics[width=11cm]{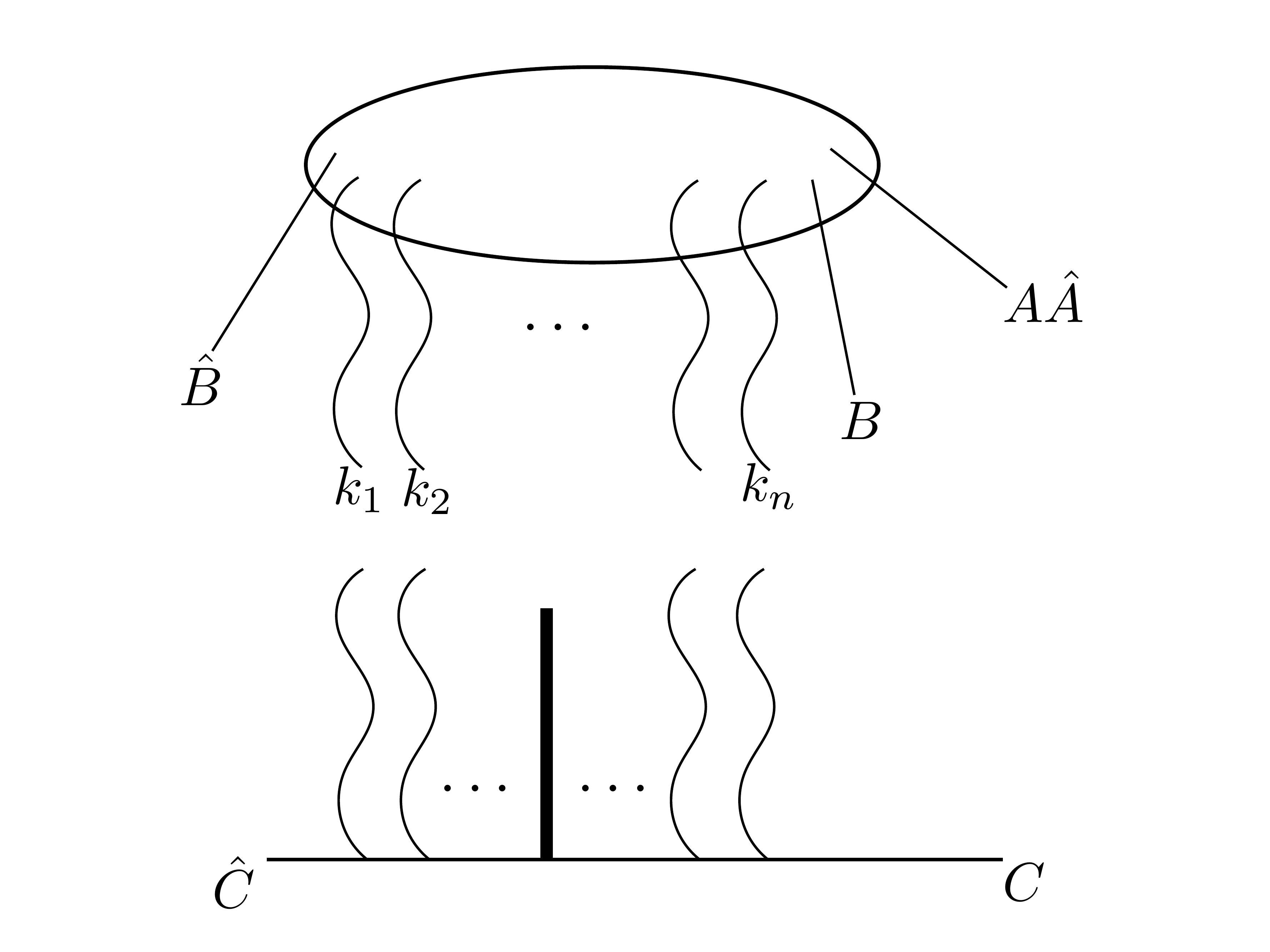}
\end{center}
\caption{The building blocks of our eikonal amplitude. The top diagram corresponds to (\ref{Eikprod}), and the diagram on the bottom corresponds to (\ref{otherside}). The thick line represents the hard Pomeron, and we must sum over crossed diagrams in the bottom half. We glue the two parts and integrate over all momenta to obtain the $(n-1)$-loop contribution to the eikonal.}
\label{buildingblock}
\end{figure}

To start, we consider the sub-diagram with external Pomeron vertex operators for $(C\hat C)_i$, $i=1,\dots n$ as in figure \ref{buildingblock}. Effectively, we have cut the $(n-1)$-loop amplitude, treating the exchanged Pomerons as external particles which will then be glued to obtain the loop result. This tree-level diagram gives
\bea\label{FirststepA}
&& \prod_r \frac{(\alpha' s_{CD})^{2+\alpha' t_{C\hat C, r}/4}}{\alpha' t_{C\hat C, r}}\int d^2 z_{\hat A} |z_{\hat{A}}|^{K_{\hat B \hat A}/2 } |1-z_{\hat A}|^{K_{A\hat{A}}/2} \int\prod_i  d^2 z_{(C\hat C)_i}\prod_{l\ne m}|z_{(C\hat C)_l}-z_{(C\hat C)_m}|^{K_{(C\hat C)_l (C\hat C)_m}/2}
\nonumber\\& & ~~~~~~~~~~~~~~~~~
\prod_j  
 |z_{(C\hat{C})_j}|^{K_{\hat B (C\hat C)_j}/2-2+K_{C\hat C, j}/4}\,|1-z_{(C\hat{C})_j}|^{K_{(C\hat{C})_j A}/2}|z_{\hat A}-z_{(C\hat{C})_j }|^{K_{(C\hat{C})_j \hat A}/2} \nonumber\\
 %\times K_{polarization}(\{z\}, \{k\})
\eea 
Once again, the Pomeron operator for the string $A\hat A$ contracts with the $\hat B$ vertex operator, leading the same Regge amplitude $\mathcal{A}_{A+B\rightarrow \hat{A}+D}$ for the auxiliary process (alternatively, we can look as the saddle point equations for $z_{\hat{A}}$ to find that $z_{\hat{A}} \rightarrow 1$).  The Pomerons $(C\hat C)_i$ will behave similarly, and since the kinematic variables $K_{(C\hat C)_j (C\hat C)_k}$ are much smaller than the other Mandelstam invariants involving the momenta $k_j = k_{(C\hat C)_j}$, we can ignore Pomeron-Pomeron interactions. This is a similar statement to keeping only the ladder diagrams (so, ignoring H-diagrams and the like) in a field theory.

If we neglect these exponents $K_{(C\hat C)_j (C\hat C)_k}$, the amplitude factorizes, giving
\be\label{Eikprod}
\mathcal{A}_{A+B \rightarrow \hat{A}+D}  \prod_{i=1}^n \frac{(\alpha' s_{CD})^{2+\alpha' t_{C\hat C, i}/4}}{\alpha' t_{C\hat C, i}} F(\alpha' k_{D_i}^2/2, \eta_i)
\ee  
where $k_{D_j}^2 = (k_{\hat B} + k_j)^2$ and $ \eta_j = \alpha' k_{A\hat A}\cdot k_j$. So far, this is a tree-level contribution, which will be a building block for our Eikonal amplitude.  In the EFT limit, the factors of F reduce to propagators at (say) the massless pole, $F\sim 1/k_D^2$.  The single string diagram reproduces the product form for the sum of all diagrams with $n$ soft graviton (or gravi-reggeon) lines emitted from a hard line derived in \cite{Weinbergsoft}.   

%We would like to capture the contribution of this to the spreading, which arises from the phase of appropriate terms in $F$ as reviewed above at tree level. 

We now want to join this tree-level amplitude to the $C\to \hat C$ leg.  This is similar to, but not identical, to the stringy Eikonal amplitude (it is not identical because $D$ is off-shell). 

The $C\to \hat C$ leg is made of GR propagators, as all of the Mandelstam invariants in this leg are small.
Following the same procedure as \cite{Weinbergsoft}\cite{Kabat}, we obtain a factor

\begin{equation}\label{otherside}
\frac{1}{n!}\prod_{j=1}^{n-1} \Big(\frac{1}{-2p_{\hat{C}} \cdot k_j - i\epsilon}+ \frac{1}{-2p_{C} \cdot k_j - i\epsilon}\Big)
\end{equation}
Notice that we only have $n-1$ factors, not all $n$ momenta $k_j$ appear. We have separated out one of the Pomerons, treating the other ones as soft. To obtain this factor, our starting point is the propagator of each Pomeron

\begin{equation}
\frac{1}{(p_j-k_j)^2 + m^2 - i\epsilon}
\end{equation}
In the Eikonal approximation, we take $p$ to be nearly on-shell so $p^2 + m^2 = 0$, and $k^2$ is taken to be small. Furthermore, due to the small momentum transfer coming from the soft modes (so we can approximate $k_i \cdot k_j \simeq 0$ for all $i,j$) the products $p_j \cdot k_j$ are equal to either $p_C \cdot k_j$ or $p_{\hat{C}} \cdot k_j$, depending on whether the soft Pomeron $(C\hat{C})_j$ emerged from the $C$ or the $\hat{C}$ side of the hard Pomeron. Since we must sum over all Feynman diagrams, we sum these contributions for each soft Pomeron $j$ and then we take the product to obtain Equation \ref{otherside}.

If we further approximate $p_{\hat{C}} \cdot k_j \simeq -p_{C} \cdot k_j $, then we replace the above factors with delta functions $\sim i \pi \delta(p_C \cdot k_j)$. This ensures $k_j^+ = 0$ for $j = 1, ..., n-1$ if we choose the longitudinal directions so that $p_C = p_C^-$. Furthermore, integrating out each $k_j$ gives a $1/p_C^-$ factor. The last $k_n$ satisfies $k_n^+ = p_{C\hat{C}}^+$ by conservation of momentum, which will be of order $q^2/p_C^-$ in our setup and can be approximated as 0. 

Thus, if we glue together the tree-level amplitude \ref{Eikprod} with the $C\to \hat C$ leg, we finally get

\begin{equation}
\mathcal{A}_{(n-1)-\text{loop}} \sim \mathcal{A}_{aux}\frac{i^n g_s^{2n}}{n!} p_C^- \int dx^+d^2x_\perp e^{-i p_{C\hat{C}} \cdot x} \prod_{j=1}^{n} \int d^{d-2} k_{j,\perp} dk_j^- e^{ik_j \cdot x} \frac{s_{CD}^{2-k_{j, \perp}^2/4}}{- k_{j, \perp}^2} \frac{1}{p_C^-} F(p_{\hat{B}}\cdot k_j, \tilde{\eta}_j)
\end{equation}
where we tentatively set $\alpha' = 1$ for brevity and wrote $k_{D_j}^2 \simeq 2 p_{\hat B} \cdot k_j$. The $k_j^+ = 0$ condition allowed us to drop the $dk_j^+$ integrals and simplify $t_{C\hat{C}, i}/4 = -k_{j, \perp}^2/4$. We re-wrote the amplitude using an ``impact parameter" representation (with an additional longitudinal separation $x^+$ as opposed to the usual, exclusively transverse separation $x_{\perp}$) in order to simplify the conservation of momentum conditions and make the re-summation easier.

The longitudinal separation parameter $x^+$ obfuscates the $s$ scaling of the loop amplitude, as the spreading phenomenon \cite{lennyspreading}\cite{sixpoints}\ occurs at a range of $x^+ \sim \alpha' p_{\hat{B}}^+$. We will instead re-write $x^+ = \alpha' p_{\hat{B}}^+ u$ where $u$ is a dimensionless parameter. Similarly, we will write $k_j^- = v_j/\alpha' p_{\hat{B}}^+$. We then re-write the amplitude as

\begin{align}
\mathcal{A}_{(n-1)-\text{loop}} \sim \mathcal{A}_{aux}\frac{i^n g_s^{2n}}{n!} p_C^- p_{\hat{B}}^+ \int d^{d-2} x_{\perp} du e^{i p_{C\hat{C}, \perp} \cdot x_{\perp} + i u p_{C\hat{C}}^- p_{\hat{B}}^+} \\
\prod_{j=1}^{n} \int d^{d-2} k_{j,\perp} e^{-ik_{j,\perp} \cdot x_{\perp}}  \frac{s^{2-k_{j, \perp}^2/4}}{-k_{j, \perp}^2} \frac{1}{p_C^- p_{\hat{B}}^+} \int dv_j e^{iuv_j} F(v_j, \tilde{\eta}_j)
\end{align}
Above, we approximated $p_{\hat{B}}\cdot k_j \simeq p_{\hat{B}}^+ k_j^-$ to simplify our expression for $F(p_{\hat{B}}\cdot k, \eta_j)$, a reasonable assumption in the limit where the relative boost is large. 

We can re-write $p_C^- p_{\hat{B}}^+ \sim s$ to get the usual scaling of the Eikonal amplitude (namely, an $s^1$ factor up-front and $s^1$ for each exchanged Pomeron). Now we can define our analogue of the eikonal phase

\begin{equation}\label{chiF}
\chi(x_{\perp}, u) = \int d^{d-2} k_{\perp} e^{-ik_{\perp} \cdot x_{\perp}}\frac{g_s^2 s_{CD}^{1-k_{\perp}^2/4}}{-k_{\perp}^2} \int dv e^{iuv} F(v, \eta)
\end{equation}
Then, our $(n-1)$-loop amplitude is simply

\begin{equation}\label{loopAchi}
\mathcal{A}_{(n-1)-\text{loop}} \sim \mathcal{A}_{aux}\frac{s_{CD}}{n!} \int d^{d-2} x_{\perp} du \, e^{-i p_{C\hat{C}} \cdot x}\, (i \chi(x_{\perp}, u))^n
\end{equation}
We can immediately see that once we carry out the summation over all $n = 1, 2, ...$ we obtain an exponential series (with the $n=0$ term removed), so

\begin{equation}\label{summedchi}
\mathcal{A}_{\text{Eikonal}} \sim \mathcal{A}_{aux} \int d^{d-2} x_{\perp} du e^{-i p_{C\hat{C}} \cdot x} s_{CD} (e^{i\chi(x_{\perp}, u)} - 1)
\end{equation}
Comparing with Kabat's result \cite{Kabat} for instance, one sees that the structure is entirely analogous, just with an additional longitudinal integral.  

%{\color{red} ES:  I do not yet quite see how this reduces exactly to Kabat in the case of pure GR.  There, they approximate $p_{C\hat C}$ to be transverse, and also $\chi$ does not depend on $x^+$ in his case.  If we took that limit here, we would seem to have an infinite factor from the $\int du$.   In Kabat before doing the $x^\pm$ integrals, we have the structure
%\be\label{Kabatcase}
%const \int d^4 x e^{-i p_{C\hat C}\cdot x}\frac{\Delta(x)}{\chi(x_\perp)} (e^{i\chi(x_{\perp})} - 1)
%\ee
%and combining this with the above statements and Kabat's eqn on the bottom of page 7, $\int dx^+ dx^-\Delta(x) = const \times\chi$ gives his result with no such divergence.  Clearly the answer is something very close to this.  
%}  
%
% 

\subsection{Pole Prescription and the Eikonal Phase}

In order to make sense of our above expression, we need to determine the appropriate pole prescription. As seen in Section \ref{Comparison}, at tree-level the contributions from poles were parametrically suppressed due to our choice of wavepacket. Thus, the poles were irrelevant up to parametrically suppressed corrections. However, at higher loop orders the wavepacket is unable to suppress internal momenta from running into poles, and it is essential to implement a causal pole prescription. Following Witten's $i\epsilon$ prescription \cite{iepsilon}, we will shift each Mandelstam invariant $k_J^2 \rightarrow k_J^2 - i\epsilon$ near a pole. With our convention $k_J^2 = -2k_J^+ k_J^- + k_{J,\perp}^2$, so this means that we must shift $k_\perp^2 \rightarrow k_\perp^2 - i \epsilon$ in the pole of \ref{chiF}, as is the case in field theory.

We must also treat the poles of the function $F(\alpha' k_D^2/2,\eta)$, which as defined in Equation \ref{Fdef} is a ratio of Gamma functions. The pole structure of $F$ is complicated, and furthermore the momentum integral over $k_\perp, k^-$ will take $\alpha' k_D^2, \eta$ outside the kinematic regime we employed to see spreading at tree-level. As we will see soon, the $dv$ integral in Equation \ref{chiF} can be done for fixed $\eta$, but the $dk_\perp$ integral becomes analytically intractable.

To circumvent this issue, we will use an internal, compactified dimension in addition to our $D$ non-compact dimensions. In AdS, the internal sphere can play this role, but for convenience we will treat the internal dimension as a circle of length $R$. If we allow momentum transfer in this direction, then besides the integral over transverse directions we must include a sum over internal momenta $k_i = n_i/R$. Let us assume for convenience that $k_{A\hat{A}} = n_{A\hat A}/R$ lies entirely in the internal direction, and then $\eta_i = \alpha' k_{A\hat{A}}k_i$ decouples from the $d^{d-2} k_\perp dk^-$ integral and it only enters in the internal mode sum.

We will assume that $-\eta_i = -\alpha' n_{A\hat A} n_i/R^2 \gg 1$ for all $n < 0$, which means that we want
\begin{equation}
n_{A \hat A} \gg R^2/\alpha'
\end{equation}
For $R \sim L_{AdS}$, the above inequality indicates that we are looking at a finite-$\lambda$ effect. We will also assume that only $n < 0$ contribute to $\chi(x^+, x_\perp)$ when we have $x^+ \simeq \alpha' p^+_D$, i.e. that only terms with $-\eta \gg 1$ contribute to spreading. Terms with $n \geq 0$ were seen at tree-level to be reproduced by a convergent sum of propagators, so they are described by EFT and they should not contribute to spreading.

Under the above assumptions, the eikonal phase becomes (for $\alpha' = 1$, and $x_i$ the separation in the compact dimension)

\begin{equation}\label{ChiExp}
\chi(x_\perp, u) = \sum_{n < 0} \int d^{d-2} k_\perp e^{-in x_i/R} e^{-i k_\perp \cdot x_\perp} \frac{G_N s_{CD}^{1- k_{\perp}^2/4 - n^2/4R^2}}{-k_{\perp}^2 - n^2/R^2+i\epsilon} \int dv e^{iuv} F(v, -n_A n/R^2)
\end{equation}
We still need to assign an $i\epsilon$ prescription to poles in $v = \alpha' k_D^2/2$. For fixed non-integer $\eta$, the function $F(v, \eta)$ has two series of poles:

\begin{equation}
1 + \frac{\alpha' k_D^2}{4} + l = 0, \quad -1-\frac{\alpha' k_D^2}{4}-\frac{\eta}{2} + l = 0
\end{equation}
for non-negative integers $l\geq 0$. The first series corresponds to on-shell intermediates states $\alpha' k_D^2 = -4(l-1)$ (for $l = 0$, this is a tachyon state). This means that we must shift $\alpha' k_D^2 \rightarrow \alpha' k_D^2 - i\epsilon$ near the first series of poles. As we will see soon, it is important to make this replacement only in the vicinity of the first series of poles, and not throughout the expression \ref{ChiExp}.

For the second series of poles, we can rewrite 
\begin{equation}
-1-\frac{\alpha' k_D^2}{4}-\frac{\eta}{2} = 1 + \frac{\alpha' k_{D'}^2}{4}
\end{equation}
where $k_{D'}^2 = (k_{C\hat{C}} + k_B)^2$. This series of poles is to be expected due to the $SL(2,\mathbb{C})$ invariance of the tree-level worldsheet which ensures that the amplitude is symmetric under $B \leftrightarrow \hat{B}$ (equivalently, we can think of $k_{D'}$ as being the time-reversed version of $k_D$). Thus, in this second series of poles we shift $\alpha' k_{D'}^2 \rightarrow \alpha' k_{D'}^2 - i\epsilon$, which is the ``opposite" of what we would get if we naively set $\alpha' k_D^2 \rightarrow \alpha' k_D^2 -i \epsilon$ (which, due to $k_B \simeq -k_{\hat{B}}$, would imply a shift $\alpha' k_{D'}^2 \rightarrow \alpha' k_{D'}^2 + i\epsilon$).

As a check of this, we note that the above prescription is analogous to what we one obtains in field theory, which would directly apply to our amplitude in the regime where one can expand $F$ as a convergent series of propagators. When such a convergent sum exists (i.e. for $\eta > 0$), we can write
\begin{equation}
F = \sum_{l \geq 0} c_l(\eta) \Bigg( \frac{1}{1+\frac{\alpha'k_D^2}{4}+l-i\epsilon} +  \frac{1}{1+\frac{\alpha'k_{D'}^2}{4}+l-i\epsilon}\Bigg)
\end{equation}
Expanding out $k_D^2, k_{D'}^2$ and making the usual Eikonal assumption $k_j^2 \ll |k_{\hat{B}} \cdot k_j|, |k_B\cdot k_j|$, we can write this in the form 
\begin{equation}
F = \sum_{l \geq 0} c_l(\eta) \Bigg( \frac{1}{\frac{\alpha' k_{\hat{B}}\cdot k_j}{2}+l-i\epsilon} +  \frac{1}{\frac{\alpha' k_{B}\cdot k_j}{2}+l-i\epsilon}\Bigg)
\end{equation}
For $l = 0$, we note that this is analogous to the factor we get from each soft Pomeron in Equation \ref{otherside}. The higher $l$ factors come from excited intermediate states, and in the $\alpha' \rightarrow 0$ limit we recover the same result (pole prescription and all) as the eikonal approximation in EFT  \cite{Kabat}.

Now that we have sorted out the pole prescription, let us move on with the calculation of the eikonal phase \ref{ChiExp}. In the spreading region $0 < \alpha' k_D^2 < -2\eta$, the function $F(v, \eta)$ is written as a product of sines and a term which is (in the Stirling approximation for large arguments)

\begin{equation}
\exp \Big\lbrace -\eta \Big( \frac{v}{-\eta} \log(\frac{v}{-\eta}) + (1-\frac{v}{-\eta})\log(1- \frac{v}{-\eta})\Big)\Big\rbrace
\end{equation}
This term is exponentially suppressed in $\eta$, so if $n_A\alpha' /R^2 \gg 1$ (i.e. $-\eta$ is large for $n = -1$) then we expect the dominant contribution to come from the smallest possible value of $-\eta$, which is when $n = -1$.

Intuitively, we can understand this result as coming from the fact that the eikonal series arises from a series of tree-level scattering events, and each tree-level amplitude is proportional to

\begin{equation}
e^{-\Delta X^+/\Delta X^+_{spr}}
\end{equation}
where $\Delta X^+_{spr}$ is the spreading radius
\begin{equation}\label{SpreadingRadius}
\Delta X^+_{spr} \sim \frac{p^+_D}{k_D^2} \sim \frac{\alpha' p^+_D}{-\eta}
\end{equation}
The tree-level result is then exponentially suppressed in $\eta$ (in tree-level, this suppression manifested as the $2^{\eta}$ factor), and thus the strongest interactions come from the smallest non-zero values of $-\eta$ \footnote{If we had non-zero $k_{A\hat{A}}$ components in the non-compact dimensions, this behavior would lead to a difficult to analyze play-off between trying to take $-\eta$ small while also staying in the regime $-\eta \gg 1$ where spreading is demonstrably present at tree-level. By using a compactified dimension with discrete momenta, we could arrange for a situation where the smallest positive value of $-\eta$ is still very large.}.

By keeping only the $n = -1$ term, and assuming $\alpha/R^2 \ll 1$ we thus obtain

\begin{equation}
\chi(x_\perp, u) = \int d^{d-2} k_{\perp} e^{-i k_\perp \cdot x_\perp} \frac{G_N s_{CD}}{-k_\perp^2 - 1/R^2} \int dv e^{iuv} F(v, -n_A/R^2)
\end{equation}
The $dv$ integral is a $k_\perp$-independent quantity, so we will write it as $\tilde{F}(u, -n_A/R^2)$. The remain integral is a massive propagator in $d-2$ spatial dimensions. The exact result can be written in terms of Bessel functions, but for now we will simply note that it scales as
\begin{equation}\label{chiscaling}
\chi(x_\perp, u) \sim \frac{G_N s_{CD} e^{-x_\perp/R}}{x_\perp^{d-4}} \tilde{F}(u, -n_A/R^2)
\end{equation}
for $d > 4$, and we get a logarithm $\log(x_\perp/R)$ at $d = 4$. The Fourier transform $\tilde{F}$ can be obtained by a contour integral, but we want to first point out that it is a boost-invariant quantity, as $u = x^+/\alpha'p_{\hat{B}}^+$ if boost independent, so it won't affect the boost dependence of $\chi$.

To perform the contour integral, we consider residues from both series of poles $1/(\alpha'k_D^2 + n - i\epsilon)$ and $1/(\alpha' k_{D'}^2 + n - i\epsilon)$. The first series gives
\begin{equation}
\Theta(-x^+)\sum_{n=0}^{\infty} 4\pi^2 i {n-\frac{\eta}{2} \choose \frac{\eta}{2}}^2 e^{\frac{i n x^+}{\sqrt{2}\alpha' p^+_{\hat{B}}}} = \Theta(-x^+) _2F_1(1 - \eta/2,1 - \eta/2,1; e^{\frac{i  x^+}{\sqrt{2}\alpha' p^+_{\hat{B}}}})
\end{equation}
For the second series, we have $p^+_B \simeq -p^+_{\hat{B}}$ and thus we get
\begin{equation}
\Theta(x^+)\sum_{n=0}^{\infty} 4\pi^2 i {n-\frac{\eta}{2} \choose \frac{\eta}{2}}^2 e^{\frac{-i n x^+}{\sqrt{2}\alpha' p^+_{\hat{B}}}} = \Theta(x^+) _2F_1(1 - \eta/2,1 - \eta/2,1; e^{\frac{-i x^+}{\sqrt{2}\alpha' p^+_{\hat{B}}}})
\end{equation}
In tree-level, we saw that string spreading peaked at a longitudinal separation \ref{Xplusstar}. In the vicinity of this separation, we have $\exp\big({\frac{\pm i n x^+}{\sqrt{2}\alpha' p^+_{\hat{B}}}}\big) \simeq -1$ and  and $x^+ > 0$, so only the second series contributes. For the real part of $\tilde{F}$ we have the approximate expression
\begin{equation}\label{ApproximateReal}
\Re\tilde{F} \simeq \frac{2^{\eta/2}}{\sqrt{\frac{\pi (-\eta/2)}{2}}}\cos(\pi \frac{\eta}{4})e^{-4\eta(\frac{x^+}{\sqrt{2}\alpha' p^+_{\hat{B}}}-\pi)^2}
\end{equation}
At $x^+ = x^+_*=\sqrt{2}\pi \alpha' p^+_{\hat{B}}$, the imaginary part is zero, but it is non-zero in general. In the vicinity of $x^+ = x^+_*$, we have the approximate expression
\begin{equation}
\Im \tilde{F} \simeq \frac{-2^{\eta/2}\sqrt{-\eta}}{5}\cos(\pi \frac{1+\eta}{4})\frac{(x^+ - x^+_*)}{\sqrt{2}\alpha' p^+_{\hat{B}}}
\end{equation}
We can see that the imaginary part becomes comparable to the real part for values $x^+ - x^+_* \sim -1/\eta$, which are well-within the wavepacket width $\sim 1/\sqrt{-\eta}$ we used to demonstrate string spreading at tree level. This imaginary part is not immediately inconsistent with unitarity, as we aren't considering $2\rightarrow 2$ scattering (where, at impact parameter space, the eikonal phase is real for elastic scattering; and the imaginary part is always positive). The negative imaginary part for $x^+ > x^+_*$ may seem worrisome at first since it gives an exponentially growing contribution to the amplitude, but this contribution is suppressed by the rapidly decaying Gaussian wavepacket $e^{-(x^+ - x^+_*)^2/2\sigma^2}$. Note that the wavepacket decays faster than the growing contribution regardless of the relative boost, since the width $\sigma \sim \alpha' p^+_{\hat{B}}$ grows with boosts as well.

The wavepacket ensures that our amplitude stays controlled despite the imaginary part, and it ensures that the leading contribution comes from the region $x^+ \simeq x^+_*$. This suggests that we can use a saddle point approximation to find the leading contribution to the eikonal series, and in particular we can follow the approach of \cite{Giddings}\cite{ACV}\ to find the dominant momentum transfer as a function of impact parameter.

\subsection{Saddle Equations and Momentum Transfer}\label{SaddleSection}

If we include the longitudinal wavepacket, the eikonal amplitude can be written in the form
\begin{equation}
\int d^{d-2} x_\perp dx^+ e^{-(x^+ - x^+_*)^2/2\sigma^{+2}} e^{-ip_{C\hat{C}}^- (x^+-x^+_*)} e^{-ip_{C\hat{C}}^\perp \cdot x_\perp}( e^{i\chi(x_\perp, x^+)} - 1)
\end{equation}
where $p_{C\hat{C}}^-$ and $p_{C\hat{C}}^\perp$ are the longitudinal and transverse momentum transfer respectively. We solve for the saddle equations by extremizing the exponential terms with respect to $x_\perp, x^+$, and we obtain
\begin{equation}
\frac{-(x^+ - x^+_*)}{\sigma^{+2}} - ip^-_{C\hat{C}} + i \frac{\partial \chi}{\partial x^+} = 0
\end{equation}
\begin{equation}
- ip^\perp_{C\hat{C}} + i \frac{\partial \chi}{\partial x_\perp} = 0
\end{equation}
In the longitudinal equation, note from (\ref{chiscaling}) that the $\chi$ term is suppressed by the string coupling $g_s^2$ relative to the other terms. While in the transverse equation the relative boost enhances the $\chi$ term and allows it to overcome these suppressions, in the longitudinal equation both $p^-_{C\hat{C}}$ and the $\chi$ term are invariant if we boost the late system while holding the early system ($C, \hat C$) fixed. Thus, we can find a solution where $x^+ = x^+_*$ and the momentum transfer is imaginary but suppressed by the string coupling
\begin{equation}
p^-_{C\hat{C}} \sim i g_s^2 p_C^- 2^{\eta/2} \sqrt{-\eta}
\end{equation}
In the weak coupling limit $g_s^2 \rightarrow 0$, the longitudinal momentum transfer vanishes and we have a well-behaved saddle. Note that in this limit, the term $p^-_{C\hat{C}}(x^+ - x^+_*)$ remains large, but it is canceled by the linear $x^+ - x^+_*$ in the eikonal phase.

We can now plug in $x^+ = x^+_*$ into the transverse equation, the imaginary part of $\chi$ thus drops out and we can use the approximate Equation \ref{ApproximateReal}. The value of the eikonal phase on the saddle is thus pure real, and the momentum transfer is
\begin{equation}\label{momtransfer}
p_{C\hat{C}}^{\perp} \sim \frac{G_N s_{CD} e^{-x_\perp/R}}{x_\perp^{d-5}} \frac{2^{\eta/2}}{\sqrt{\frac{\pi (-\eta/2)}{2}}}\cos(\pi \frac{\eta}{4})
\end{equation}
in the limit $x_\perp \ll R$. In the opposite limit, replace the $x_\perp$ in the denominator with $R$. This expression is very similar to the gravitational eikonal, but it has a few differences. The first is that we have an additional suppression of $2^{\eta/2}$, and the graviton exchange has been replaced by the exchange of a Kaluza-Klein mode with mass $1/R$. A similar behavior arises in scattering of particles near AdS black holes, where the curvature achieves a similar effect with $1/R\sim 1/l_{AdS}$. 

A more significant difference from the gravitational eikonal is that the sign of the eikonal phase is not positive (even on the saddle point where the phase is real). This effect is probably related to the fact that we kept $A\hat{A}$ as a fixed momentum state (rather than writing the amplitude in terms of an impact parameter for $A\hat{A}$). The eikonal phase is positive only in impact parameter space, and in our 6-point scattering we used a mix of momentum and impact parameter space. We expect that a ``full impact parameter space" expression would be positive.

\section{The Infaller-Detector System}\label{DetectorKinematics}

\begin{figure}
\begin{center}
\includegraphics[width=0.6\textwidth]{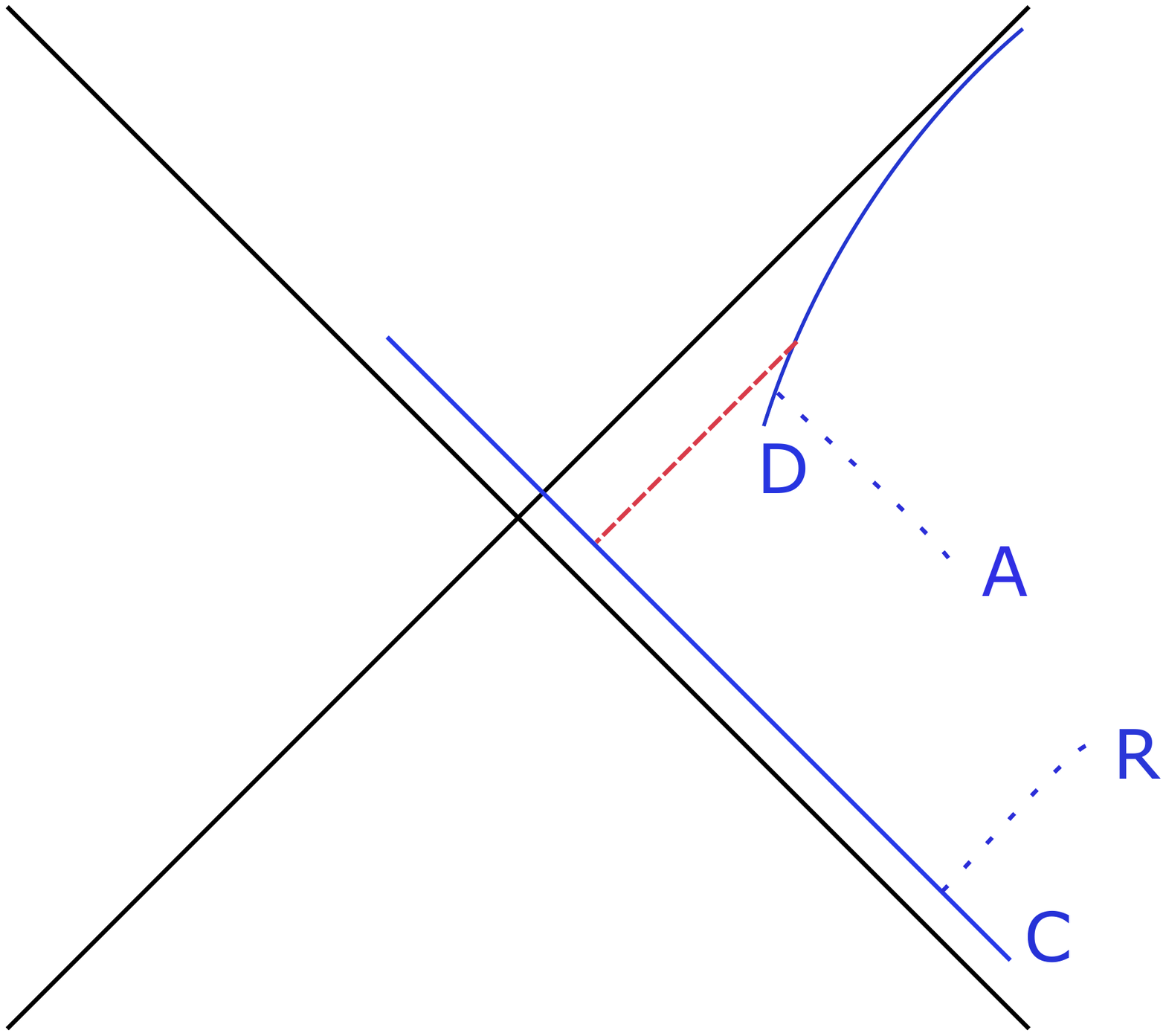}
\end{center}
\caption{An infaller $C$ is thrown in at some early time $t_0$. A time $\Delta t$ later, a detector $D$ is lowered to the near-horizon region, and it interacts with $C$ via longitudinal spreading. By measuring the change in the transverse momentum of $D$, we can gain some information about the infaller's initial state. The systems $R, A$ are reference systems which we will use in Section \ref{InfoRecovery} in order to keep track of the quantum state of the infaller-detector system.}
\label{DetectorInfallerSetup}
\end{figure}

Having established the persistence of string spreading throughout the eikonal regime, we now wish to apply our flat spacetime results to an infaller-detector setup in the near-horizon region of a Schwarzschild black hole. We will use the $C+D \rightarrow \hat{C}+\hat{D}$ sub-process (we re-label $\hat{B}$ as $\hat{D}$) to model the interaction between an infalling system and a detector held at fixed proper distance (see Figure \ref{DetectorInfallerSetup}) . Here, we will have $C$ play the role of the infaller, which we will assume is a collection of closed strings. The off-shell string $D$ plays the role of a detector that is lowered to the near-horizon region a time $\Delta t \gg r_s$ later, and which can interact with $C$ thanks to longitudinal spreading. 

The detector is assumed to be made of closed strings and held at a fixed
position $r=R$ in the Schwarzschild radial coordinate $r$.  
This corresponds to a fixed proper distance $L_{pr}=2\sqrt{r_s(R-r_s)}$ from the horizon along constant Schwarzschild time slices. From a scattering-matrix perspective, the act of holding $D$ in place amounts to repeated scattering with closed strings that impart the necessary force for $D$ to be stationary. The auxiliary strings in the 6-point amplitude thus play the role of the apparatus that holds the detector string $D$ near the black hole.

In this section, our goal is to show that by keeping track of the initial and final state of the detector $D$ we can obtain information about the kinematics of the infaller $C$. The transverse momentum $q_\perp$ imparted upon the detector is sensitively dependent on the center-of-mass energy $s_{CD}$ and the impact parameter $x_\perp$. Aside from the longitudinally non-local interaction, this process is rather prosaic; we are merely probing $C$ as it falls into the black hole, and we wish to show that our ``experiment" is sensitive to the initial data of $C$. If string spreading was compatible with an absorption effect where the detector can cleanly capture the infaller, then there wouldn't be any need for further calculation. However, since we have only demonstrated spreading for elastic processes, it's not entirely clear how much information about an infaller's state we can recover. A more careful analysis will be left for Section \ref{InfoRecovery}; in this section we will give a quick estimate of the \textit{classical} information that we can recover. We will later see that this estimate will be useful in estimating the \textit{quantum} information we can recover in a variant of the Hayden-Preskill protocol.

Suppose we choose $C$ among an ensemble of Gaussian wavepackets with fixed energy $E_C$ and transverse widths $\Delta x_\perp, \Delta q_\perp$. Our goal is to determine the initial state of $C$ by measuring the momentum transfer on $D$. When a complete determination is not possible, we will want to maximize the amount of information we obtain about the initial state of $C$, measured in terms of the Shannon entropy. 

The maximum possible amount of classical information we could obtain by a scattering experiment $C+D \rightarrow \C + \D$ is equal to the logarithm of the number $|\hat{D}|$ of distinguishable states that the detector $D$ can scatter into as we vary the initial state of $C$. This number is determined by two main factors: the strength of the scattering and the sensitivity of the detector. We will estimate both, and find that we can recover a significant amount of classical information. One important caveat is that we can only recover information that is encoded in $C$'s kinematics; the scattering experiment would have to be sensitive to $C$'s internal quantum numbers to recover any further information.

We will then generalize our considerations from $2\rightarrow 2$ to processes with more $C$ and $D$ strings, and find backreaction constraints on the number of strings $N_C, N_D$ that we can consider. These results will be useful in Section \ref{InfoRecovery} in order to estimate our capacity for quantum information recovery.

\subsection{Detector Kinematics}\label{DetectorKinematicsSection}

In order to interpret the physical results of the eikonalized interaction (i.e. momentum transfer and time-delay), it is important to understand the quantity $\alpha' k_D^2$ which appears in the 6-point amplitude. There, $\alpha' k_D^2$ is positive as $D$ is off-shell, while here we will take $D$ to be a physical object with positive mass $-\alpha' k_D^2 > 0$. In \cite{BHpaper}, a treatment of the quantized string in lightcone gauge gave the spreading estimate
\begin{equation}\label{SpreadingEstimateBHpaper}
X^+_{spr} \sim \frac{p_D^+}{k_{D, \perp}^2 + m^2_D}, \text{ when } k_{D, \perp}^2 + m^2_D \gg 1/\alpha'
\end{equation}
For $X^+ \gtrsim X^+_{spr}$, the effective ``mass distribution" of the string decays exponentially as $e^{-X^+/X^+_{spr}}$, and one expects an analogous exponential suppression to our interaction. In Equation \ref{SpreadingRadius}, we argued that the $2^{\eta/2} \simeq 2^{-\alpha' k_D^2}$ term we found in our amplitude was this exponential suppression, and thus the ``spreading radius" of our amplitude is
\begin{equation}\label{SpreadingEstimateAmplitude}
X^+_{spr} \sim \frac{p^+_D}{k_D^2}
\end{equation}
This is not to be confused with the value $X^+_* \sim \alpha' p^+_D$ where our amplitude peaked; the ratio $X^+_*/X^+_{spr} \sim \alpha' k_D^2$ is much larger than 1 and is why we got a suppression $2^{-\alpha' k_D^2}$ in our amplitude.

Based on the above similarities, we will identify the off-shell mass $k_D^2$ with the quantity $k_{D,\perp}^2 + m_D^2$ of the hovering detector. As an aside, note that both Equations \ref{SpreadingEstimateBHpaper} and \ref{SpreadingEstimateAmplitude} were derived from saddle points that required $\alpha'(k_{D,\perp}^2 + m_D^2) \gg 1$ and $\alpha' k_D^2 \gg 1$ respectively, giving an additional piece of evidence for the identification. 

Modeling the interaction of $D$ with the infaller $C$ with our amplitude from Section \ref{LoopCorrections}, we find using (\ref{momtransfer}) that the momentum transfer is almost purely transverse and proportional to
\begin{equation}\label{DeflectionFormula}
\Delta q_{\perp} \sim \frac{G_N s_{CD} e^{-x_\perp/R}}{x^{d-5}_{\perp}} 2^{-\alpha' (k_{D, \perp}^2+m_D^2)}
\end{equation}
At large $G_N s$, the deflection $\Delta q_{\perp}$ will have a significant impact on the final state $\hat{D}$, which will be sensitive to the exact parameters $G_N s_{CD}, x_{\perp}$. Thus, if we had knowledge of the initial state $D$ and the final state $\hat{D}$, we could infer information about the kinematics of $C$. Intuitively, the larger we can make $G_N s_{CD}$, the stronger the deflection and the ``signal" that we get.

However, if we take $G_N s_{CD}$ to be too large, we can run into issues of backreaction. One limiting factor is the regime $x_{\perp}^{D-3} \sim G_N \sqrt{s_{CD}}$ where corrections to the eikonal become important and subsidiary black holes may form. A more severe limitation is that the Shapiro time-delay 
\begin{equation}
\Delta X^-_D \sim \frac{\partial \chi}{\partial p^+_D}
\end{equation}
can push the detector behind the black hole horizon if $\Delta X^-_D > X^-_D$. Note that in our amplitude, we did not find the eikonal phase to be positive and with the simple form we found in (\ref{chiscaling}) it naively looks as though this time-delay could in principle be negative. However, we expect this to be an artifact of the fact that we mixed impact-parameter and momentum spaces in our calculation (by keeping $A\hat{A}$ at fixed momentum), and a more complete calculation would give a positive phase with the same parametric dependence on $\eta \simeq -\alpha' k_D^2$ and $G_N s$. 

The exact form of the amplitude is unimportant if we wish to find the maximum momentum transfer. The constraint $\Delta X^-_D < X^-_D$ gives

\begin{equation}
\chi = \frac{\partial \chi}{\partial p^+_D} p^+_D \lesssim X^-_D p^+_D
\end{equation}
The first equality follows from the linearity of $\chi$ with respect to $p^+_D$, and the second follows from the Shapiro time-delay formula. Parametrically, the transverse momentum transfer is $\chi/x_\perp$ (or $\chi/R$ for $x_\perp \gtrsim R$, where $R$ is the internal dimension we used in our amplitude), so we find that the maximum momentum transfer is
\begin{equation}
\Delta q_\perp \sim \frac{X^-_D p^+_D}{x_\perp}
\end{equation}
We can rewrite this in terms of the proper distance and energy of $D$ as
\begin{equation}
\Delta q_\perp \sim \frac{L_{pr} E_{D,pr}}{x_\perp}
\end{equation}
If we hold our detector at fixed $L_{pr}$, then our detector's proper energy is $E_{D,pr} \sim m_D$. For any fixed $m_D$, in order to maximize the momentum transfer (and thus get a clear signal from the infaller) we then seek to maximize the proper distance. However, we cannot increase $L_{pr}$ arbitrarily much without ruining the interaction; for our amplitude to be a valid description of the interaction we require

\begin{equation}
X^+_* \sim \alpha' p^+_D > X^+_D - X^+_C \simeq X^+_D
\end{equation}
where we used the fact that at late times $X^+_D \gg X^+_C$. This can be equivalently be written in terms of $L_{pr}, E_D$ as
\begin{equation}\label{KinematicConstraints}
E_{D,pr} > \frac{L_{pr}}{\alpha'}
\end{equation}
To maximize the deflection $\Delta q_{\perp}$, we must then take $L_{pr}$ to be as large as our available detector energies $E_{D,pr}$ will allow. Before we go on further, let us comment that the local energies $E_{D,pr}$ are blueshifted compared to the energies $E_D$ measured by an asymptotic observer at $r \gg r_s$ as
\begin{equation}
E_{D,pr} \sim E_D \frac{r_s}{L_{pr}}
\end{equation}
In terms of the asymptotic energies, the required detector kinematics are
\begin{equation}\label{EDIneq}
E_D > \frac{L_{pr}^2}{r_s}(k_{D,\perp}^2 + m_D^2) \gtrsim \frac{L_{pr}^2}{r_s\alpha'}
\end{equation}
In two extremes, we can take $L_{pr} \sim l_s$ in which case $E_D \sim 1/r_s$ is sufficient, or we could take $L_{pr} \sim r_s$ which would require $E_D > r_s/\alpha' \sim g_s^2 M_{BH}$. This constraint is reminiscent of a constraint on the detector mass found in \cite{BHpaper}. 

Note that while $L_{pr} \sim l_s$ is expected of the old stretched horizon picture, we see that higher-energy detectors can pick information about the infaller at a parametrically larger distance from the horizon. The fact that it is possible to recover an infaller's information at a large distance $L_{pr} \sim O(r_s)$ from the horizon implies that black hole information, and the associated EFT violation, is highly non-local over macroscopic length scales (as long as it is probed with the right equipment). Still, we need to remain in the near-horizon region for flat space amplitudes to be a good approximation, so even if we take $L_{pr}$ parametrically similar to $r_s$ we still still need to keep $L_{pr} \ll r_s$.  

If we optimize the kinematics of our detector by taking $L_{pr}$ as large as possible, and the relative boost to be as large as backreaction allows, then the momentum transfer to $D$ is
\begin{equation}
q_\perp \sim \frac{L_{pr}^2}{x_\perp \alpha'} \sim E_D \frac{r_s}{x_\perp}
\end{equation}
As long as $E_D \gg 1/r_s$, the transverse momentum transfer is large compared to the background scale, and we expect a noticeable ``kick" to the detector. If we had a momentum detector with infinite precision, then we would be able to detect the most minute differences in the momentum transfer and easily recover information about the infaller. However, the uncertainty principle combined with the finite size of the black hole horizon forbids such precision, and a careful accounting of the detector precision is necessary to estimate our capacity to recover information.

\subsection{Detector Resolution and Classical Information Recovery}\label{DetectorResolutionSection}

Suppose that we wish to encode a message in the state of the infaller $C$. The best way to do this is to associate a string of bits to a set of orthogonal states of $C$. If we then wish to recover the classical message using the $C+D$ interaction, we want to ensure that any two different states of $C$ will affect $D$ in a different manner. If this were the case, then as long as we knew the cypher that maps bit strings to states of $C$, then we could measure the state of $D$ and recover the bit string.

Of course, for this to be possible we must ensure that the detector has enough states to encode all bit strings of interest. When this is not the case, then we cannot perfectly recover the message, but we can recover some degree of information. By making a measurement on $D$, one could reduce their uncertainty about the state of $C$ (or equivalently, about the encoded message). The maximum amount of information that we can recover from the infaller is $\log_2|\hat{D}|$, where $|\hat{D}|$ is the number of distinguishable states that the detector can scatter into (assuming a given initial state). That is the shortest bit string that can fully describe the state of the detector; we can't expect to get any more information than that.

In order to make our argument as simple as possible, it will be convenient to choose a basis of states whose dynamics are approximately classical. Suppose that we write such a basis $|C_i\rangle$ for the infaller and a basis $|D_j\rangle$ for the detector. Then, we want the interaction $C+D \rightarrow \C + \D$ to take the form
\begin{equation}\label{ClassicalScattering}
|C_i\rangle |D_j\rangle \rightarrow |\C_{ij}\rangle |\D_{ji}\rangle
\end{equation}
where the outgoing states are still part of the same basis. If we were to use momentum eigenstates for example, this would not be the case; the interaction would generate superposition of momentum eigenstates. If we were to work in position space instead, then while there would be no generation of superpositions, the interaction would only generate an eikonal phase-factor. Thus, it would be hard to talk about distinguishability in position space.

We will instead work with states whose transverse position dependence is described by a Gaussian wavepacket with a narrow width $\Delta x_\perp \sim \epsilon r_s \ll r_s$. The typical impact parameter for interactions will be $x_\perp \sim r_s \gg \Delta x_\perp$, so we can approximately treat the interaction as happening at exact impact parameter $x_\perp$, leading to a transverse momentum transfer $q_\perp \sim \frac{\partial \chi}{\partial x_\perp}$. The uncertainty in this estimate is 
\begin{equation}
\frac{\partial q_\perp(x_\perp)}{\partial x_\perp} \Delta x_\perp \sim q_\perp \frac{\Delta x_\perp}{x_\perp} \sim q_\perp \epsilon
\end{equation}
When the above uncertainty is larger than the width $\Delta q_\perp$ of our wavepackets in momentum space, the above estimate ensures that the $C+D$ interaction won't take the exact form \ref{ClassicalScattering}, but instead it generates superpositions of our basis states $|\C_a\rangle |\D_b\rangle$. When $\Delta q_\perp$ is larger, these superpositions can be ignored, and Equation \ref{ClassicalScattering} holds to a good approximation. We thus need $q_\perp \epsilon < \Delta q_\perp$. However, if we take $\Delta q_\perp$ to be unnecessarily large, then we will lose the ability to distinguish the effect of the $C+D$ interaction on $D$, which amounts to a change in momentum. Thus, we will take $\Delta q_\perp \sim q_\perp \epsilon$, perhaps up to some numerical prefactors.

Now, if we wish to make the ``signal" of the $C+D$ interaction as large as possible, we must maximize the momentum transfer $q_\perp$. Following the results of the previous section, we can do so by taking the energy of the detector to be $E_D \sim r_s/\alpha'$, and thus the momentum transfer will be
\begin{equation}
q_\perp \sim E_D \frac{r_s}{x_\perp} \sim \frac{r_s}{\alpha'}
\end{equation}
With this value of the momentum transfer we see that our Gaussian wavepackets must have
\begin{equation}
\Delta x_\perp \sim l_s, \quad \Delta q_\perp \sim \frac{1}{l_s}
\end{equation}
in order to ensure that Equation \ref{ClassicalScattering} can hold. With these parameters, we can see that the relevant uncertainties are $\Delta x_\perp/x_\perp, \Delta q_\perp/q_\perp \sim l_s/r_s$ both go to zero as we take $l_s/r_s \rightarrow 0$, justifying the use of a semiclassical limit.  Here we neglect the logarithmic effect of transverse spreading \cite{lennyspreading}.  

We can build an approximate basis from these wavepackets by varying their peak positions and momenta. In order to prevent overcounting, we can choose to ensure that wavepackets with similar positions/momenta will have their peak positions differing by at least $10\Delta x_\perp$ and their peak momenta by at least $10\Delta q_\perp$ (the factor of $10$ is arbitrary and not particularly important, as we will see soon). Of course, this is not sufficient to guarantee that they will have zero overlap, just that their overlap will be a very small number. This means that we would have an overcomplete set of wavepackets, but such an overcounting can only possibly change $\log_2|D|$ by an $O(1)$ number, which is subleading in the $l_s/r_s \rightarrow 0$ limit. We will be dropping any such subleading terms in the rest of the paper, and so for simplicity we can assume that we have a family of orthogonal wavepackets whose ``spacings" in position and momentum space are $\Delta x_\perp, \Delta q_\perp$ respectively. From now on, we will refer to this collection of wavepackets as the semiclassical basis.

Now that we have established a semiclassical limit, let us go back to the issue of classical information recovery. Suppose that we choose to encode a message in the transverse kinematics of $C$ by assigning a string bit to each element $|C_i\rangle$ in the semiclassical basis. Let's take the detector to be in a fixed state $|D_j\rangle$, also in the semiclassical basis. If we change the position of $|C_i\rangle$, the impact parameter will change by at least $l_s$ and thus the momentum transfer will change by at least $\frac{\partial q_\perp(x_\perp)}{\partial x_\perp} l_s \sim 1/l_s$. This means that any measurable change in the position of $|C_i\rangle$ will lead to a measurable difference in the momentum of the detector's final state $|\D_{ji}\rangle$.

Thus, up to effects that are subleading in the $l_s/r_s\rightarrow 0$ limit, the detector can perfectly distinguish between the positions of any two elements in the semiclassical basis of $C$. The amount of classical information we can recover is then equal to the logarithm of the number of distinguishable momenta that are available to the detector. For any fixed energy $E_D$, the number of states $|D|$ available to the detector system is given by the volume of this semiclassical phase-space in units of the resolution
\begin{equation}
|D| \sim \frac{A_{BH}}{\Delta x_\perp^{d-2}} \big(\frac{E_D}{\Delta q_\perp})^{d-2} \sim \big(\frac{r_s}{l_s}\big)^{2(d-2)}
\end{equation}
However, these states can be labeled by their position and momentum $(x_i, q_i)$. Initially only the momentum is affected by the $C+D$ interaction, so we can only use the momenta to ``record" the state of $C$. Thus, the number of distinguishable final states $|\D|$ available to the detector are given by

\begin{equation}
|\D| \sim \big(\frac{E_D}{q_{res}}\big)^{d-2} \sim \sqrt{|D|}
\end{equation}
Any measuarable difference in the position of $C$ will lead to a measurable difference in the state of $\D$, so knowing the final state of the detector will give us
\begin{equation}
\log_2|\hat{D}| = (d-2) \log (\frac{r_s}{l_s})
\end{equation}
bits of information about the state of the infaller.

Now, does this mean that if $|\D| > |C|$ that we can fully recover all information encoded in the infaller $C$? The method we just described does not accomplish that; it only immediately recovers information that is encoded in the positions $x_i$ of the semiclassical basis elements $|C_i\rangle$. The transverse momenta $q_i$ do not affect the interaction nearly as strongly,  although they do at subleading orders in the string coupling \footnote{The transverse momenta enter the string amplitude through the center-of-mass energy $s = -p^-_C p^+_D + q^\perp_C \cdot q^\perp_D$. In our kinematics, there is a relative boost of up to order $1/g_s^2$ which enhances the longitudinal momenta, but it does not affect the transverse momenta. Thus, even when we have $q_\perp \sim r_s/\alpha'$, the contribution to $s$ is still smaller by a factor of $g_s^2$ compared to the longitudinal contribution.  Similarly, details of $C$'s excitation level and spin appear in the amplitude in a way that is not amplified by the relative boost.}, and thus we cannot as easily distinguish between states $|C_i\rangle$ that have the same $x_i$ but different $q_i$. Similarly, we cannot -- at least not as easily -- distinguish between states that have information recovered in internal quantum numbers.  It would be interesting to develop optimal estimation methods for detecting the small signal in these features of $C$, something that we will leave for future work.  

Still, even given these caveats we have demonstrated a rather concrete demonstration of classical information recovery using longitudinal string spreading. The recovery of quantum information on the other hand is fundamentally different, and it cannot be immediately deduced from the above considerations. One crucial difference is that quantum information cannot be copied, and thus the infaller must ``forget" its initial state for the detector to fully recover quantum information. We will study the issue of quantum information recovery in Section \ref{InfoRecovery}. For the rest of this section, we will address some preliminary issues that will be useful to optimize the ``performance" of our detector setup.

\subsection{Multi-String Infallers and Detectors}\label{MultiString}

So far we have considered a setup where there is a single infalling string $C$, and a single detector string $D$. However, more generally we could have a large number of infallers (which we will collectively denote as $C$), and we could consider a detector $D$ made from a large number of strings. We will write each state in $C$ as $|C_I\rangle = |C_{i_1}\rangle ... |C_{i_{N_C}}\rangle$, where each $|C_{i_k}\rangle$ is in the semiclassical basis and we take $N_C$ to be a fixed number. In order to avoid the issue of quantum statistics, we will take the energies of all strings to be distinguishable, or alternatively we can take $N_C$ to be much smaller than the number of available states per string. This is reminiscent of the dilute gas limit, where both Bose and Fermi statistics approach the Boltzmann distribution. Similarly, we will write a basis for $D$ in the form $|D_J\rangle = |D_{j_1}\rangle... |D_{j_{N_D}}\rangle$, and once again ensuring that the various constituent strings are essentially distinguishable. Using multiple strings will vastly expand the available phase-space of the detector, and we expect that the information recovery capacity will increase as $\log|D^{max}| \propto N_D$.

To deal with such scattering processes, we will generalize our $2 \rightarrow 2$ amplitude $C+D \rightarrow \hat{C}+\hat{D}$ to more general $N_C + N_D \rightarrow N_C + N_D$ amplitudes in the eikonal regime. The leading contribution comes from ladder-diagrams with endpoints connecting the various components of $C$ and $D$. These diagrams are all independent from each other, and thus we can write the amplitude in position space as
\begin{equation}
\mathcal{A}(\lbrace s_{ij}, x_{ij}^{\perp}, x_{ij}^+ \rbrace) \sim \prod_{i \in C ,j \in D} s_{ij} (e^{i\chi(s_{ij},  x_{ij, \perp}, x_{ij}^+ \rbrace)}-1)
\end{equation}
where $i, j$ run over all strings in $C, D$ respectively, and $\lbrace s_{ij}, x_{ij}^{\perp}, x_{ij}^+ \rbrace$ are the kinematic parameters of the $C_i + D_j$ scattering event. For simplicity, we can imagine that the infallers $C$ fall into the black hole around the same time, but we can tune their transverse kinematics and energies to encode a large amount of information. After around a scrambling time, the center of mass energy between the strings $C$ and the detectors $D$ gets large, and the detector strings $D_j$ undergo a strong deflection that is given by 
\begin{equation}\label{MultipleDeflections}
q_{\perp, j} \sim \sum_{i \in C} q_\perp(s_{ij}, x_{ij}) \hat{x}_{ij}
\end{equation}
where $\hat{x}_{ij}$ is the direction of transverse separation between $C_i$ and $D_j$, and $q_\perp(s_{ij}, x_{ij})$ is the momentum transfer for the given kinematics. Because the directions $\hat{x}_{ij}$ can differ, the deflection that $D_j$ undergoes is less than what it would if it interacted with a single string that had the collected energy of all the $C_i$. However, the Shapiro time-delay

\begin{equation}
\Delta X^-_D \sim \sum_{i \in C} \Delta X^-(s_{ij}, x_{ij})
\end{equation}
would be the same regardless (as we simply have a sum of positive numbers, there is no possible cancellation). If each $s_{ij}$ is of the same order of magnitude and for a general scattering event we have $x_{ij}^\perp \sim r_s$, then we can apply the same rationale as in Section \ref{DetectorKinematicsSection} to get the estimate
\begin{equation}
\Delta q_{\perp}(s_{ij}, x_{ij}) \sim \frac{L_{pr} E_{D_j, pr}}{x_{ij}}
\end{equation}
Now, suppose that we take the infalling strings of $C$ to have positions independently and randomly chosen from some distribution\footnote{For example, we could take $C$ to be made of strings with fixed energies and uniformly and independently random positions. Our requirement of randomness is one of convenience to understand the behavior of ``generic infallers", rather than special states where all the deflections are in the same direction and enhance each other. In any case, such an enhancement would only make the signal clearer.}, then by the central limit theorem the total deflection $q_{\perp,j}$ would follow a Gaussian distribution with mean 0 and variance

\begin{equation}\label{RMStransfer}
(\Delta q_{\perp, j})_{RMS} \sim \frac{1}{\sqrt{N_C}} \frac{L_{pr} E_{D_j, pr}}{r_s} \sim \frac{1}{\sqrt{N_C}}\frac{r_s}{\alpha'}
\end{equation}
When we average over all $I$, this deflection will map each initial state $|D_J\rangle$ to a mixed state that covers a large region in momentum-space (the positions of $|D_J\rangle$ are of course left unchanged) with radius $r_s/\alpha' \sqrt{N_C}$. If we want a parametric estimate of the entropy of this mixed state (dropping any factors that are subleading in the semiclassical limit), we can simply compute the volume in this region to obtain 
\begin{equation}\label{LogDformula}
\log|\hat{D}| \sim N_D (d-2) \Big( \log(\frac{r_s}{l_s}) - \log(\sqrt{N_C})\Big)
\end{equation}
If we take $N_C$ to be a large but $O(1)$ number (compared to $r_s/l_s$), then the $\log(\sqrt{N_C})$ factor is negligible. Compared to the $2\rightarrow 2$ setting, all that we have is an enhancement of the capacity of the detector to recover classical information by a factor of $N_D$. 
Barring issues of backreaction (which we will investigate in the next subsection), the detector's ability to recover information about the infaller has gone almost unimpeded. As before, we can most easily recover information encoded in the positions of the strings that make up the infaller $C$, although with large $N_D$ it may prove possible to obtain a statistically significant detection of $C$'s transverse momenta (on which the amplitude depends weakly in our kinematic regime).

\subsection{Backreaction Constraints}\label{BackreactionSection}

In this section, we will briefly investigate backreaction constraints on the infaller-detector system. There are two constraints that we need to consider: (i) Whether the detector backreacts on the geometry, causing the black hole horizon to swell and consume it, and (ii) Whether the $C+D$ interaction goes beyond the eikonal regime of validity and into the ``black hole formation" regime. 

The second constraint is strictly less restrictive than the first combined with the Shapiro time-delay being $\Delta X^-_D < X^-_D$. For the first consideration, what we need for the detector to avoid being consumed by the black hole is (we work in 4 spacetime dimensions for simplicity)

\begin{equation}\label{NDIneq}
G_N N_D E_D < L_{pr}
\end{equation}
where recall that $E_D$ is the (asymptotic) energy of each string that makes up the detector. However, we also have $E_D > L_{pr}/\alpha'$, so we have
\begin{equation}
G_N N_D < \frac{1}{\alpha'} 
\end{equation}
This places a constraint $N_D < Vol/g_s^2$ where $Vol$ is the volume of the extra dimensions in string units. Note however that this is a limit to how many detectors we can have in place at the same time, there is nothing stopping us from lowering detectors at different times to avoid backreaction. Of course, we will still want to lower each detector $D_j$ at the ``optimal" time when $G_N \sum_i s_{ij}$ is as large as the Shapiro time-delay allows.

\section{Quantifying Information Recovery}\label{InfoRecovery}

In the previous section, we examined the kinematics of our infaller-detector system and we made some semiclassical estimates about classical information recovery. We will now examine the recovery of an quantum information by performing a variant of the Hayden-Preskill protocol \cite{HaydenPreskill}.  For simplicity of notation, we will work at $d = 4$, but our considerations naturally generalize to $d > 4$. In our argument, we will make use of four different systems: the infalling strings $C$, the detector $D$, and their respective reference systems $R, A$.

The reference system $R$ encodes the state of the infalling system $C$, so we will assume that they start in an entangled state
\begin{equation}\label{CRstate}
\sum_I c_I |C_I\rangle |R_I\rangle
\end{equation}
We will use upper-case indices when referring to the multi-particle states $|C_I\rangle = |C_{i_1}\rangle |C_{i_2}\rangle ... |C_{i_n}\rangle$, and lower-case indices to refer to individual strings if necessary. As in the previous section, we will use the semiclassical basis (see Section \ref{DetectorResolutionSection} for more details). We will assume that the different states $|C_I\rangle$ are distinguished purely by their kinematics, and we will also assume that the individual strings $C_i$ have fixed, distinct energies. This is not purely a matter of convenience; our protocol can only decode information encoded in transverse kinematics, and this fixed energy infaller is the simplest scenario in which quantum information recovery can be concretely demonstrated. The distinct energies provide a simple way to ensure that classical statistics are obeyed. In this simple example, we can take $c_I = e^{-S_C/2}$ for all $I$, i.e. the constituent strings have fixed energies but completely random transverse positions and momenta.

Similarly, we will take our detector $D$ to be entangled with a reference system $A$
\begin{equation}\label{DAstate}
\sum_J d_J |D_J\rangle |A_J\rangle
\end{equation}
The detector $D$ will be made of multiple strings lowered near the horizon to a distance $L_{pr}$ by an apparatus which we will take to be part of the reference system $A$. We allow the detector strings $D$ to stay near the horizon for some time $\Delta t \sim r_s$, where they will interact with the infaller $C$ via longitudinal string spreading. Then, we remove the detectors from the near-horizon region via the same apparatus.

The reference systems $R, A$ do not interact with $C, D$ in any manner during the $C+D$ interaction; they are simply there to encode the information of the infaller/detector system. We will quantify the recovery of quantum information about the infaller by measuring the mutual information
\begin{equation}
I(\D A: R) = S_{\D A} + S_R - S_{\D A R}
\end{equation}
where $\D$ denotes the detector system after its interaction with $C$. The reference system $R$ acts as a record of the infaller's initial quantum state, and thus the mutual information acts as a measure of correlation between the final state of $\D A$ and the initial state of $C$.  For two subsystems with $N$ maximally entangled qbits, the mutual information is $2N\log 2$.  Roughly speaking, it counts (twice) the number of EPR pairs between two systems, so we will say that we have acquired $\frac{1}{2}I(\D A:R)$ qubits of information about the state of $C$. We will show that in the semiclassical limit $l_s/r_s \rightarrow 0$, the mutual information gets contributions at least as large as
\begin{equation}\label{MutualInfoResult}
I(\D A: R) \gtrsim \min \lbrace S_C, S_D \rbrace
\end{equation} 
As in Section \ref{DetectorResolutionSection}, this lower bound is ``half" of the information encoded in the transverse kinematics, which arises from the strongest dependence on infaller kinematics in the amplitude. To keep our computation straightforward, we will make a few simplifying assumptions about the detector. Unlike the restrictions on the infaller, which may imply some physical limitations on our recovery protocol, we are free to design our detector whichever way we wish (as long as we stay within the backreaction constraints of Section \ref{BackreactionSection}).

First, we will assume that all states $|D_J\rangle$ have the same $N_D$, and they are made of strings with fixed, distinct energies $E_1, E_2, ..., E_{N_D} \sim r_s/\alpha'$. This energy allows our detector system to interact with the infaller while sitting at a distance $L_{pr} \sim r_s$ from the horizon. Following the results of Section \ref{DetectorResolutionSection}, this will give us the maximum possible resolution of the infaller kinematics. 

The only difference between the different states $|D_J\rangle$ is once again the transverse kinematics of their constituent strings: each state $|D_J\rangle$ consists of $N_D$ strings with fixed energies and variable transverse coordinates and momenta. As we did with the infaller $C$, we will use the semiclassical basis of Section \ref{DetectorResolutionSection} to describe the transverse kinematics. We will take the density matrix $\rho_D$ to be maximally mixed in the transverse kinematics of the strings that make up $D$, i.e. every transverse location and momentum direction is equally likely so $d_J = e^{-S_D/2}$ in Equation \ref{DAstate}.

In the limit $l_s/r_s \rightarrow 0$, the semiclassical basis allows us to estimate the entropy of $D$ by counting the volume of the phase-space available to $D$ in units of $\Delta x_\perp \Delta q_\perp$. The entanglement entropy of $D$, which also gives the ``information capacity" of our detector, is then given by
\begin{equation}\label{Dentropy}
S_D = N_D \log\Big(\frac{A_{BH}}{\Delta x_\perp^2} \frac{E_D^2}{\Delta q_\perp^2}\Big) \simeq N_D \log\Big(\frac{r_s^4}{\alpha'^2}\Big)
\end{equation}
The $A_{BH}$ term, i.e. the black hole area, arises from the different transverse locations available to our detector, while $E_D$ gives the available transverse momenta.

Once systems $C$ and $D$ interact via longitudinal spreading, we will denote the resulting systems as $\hat{C}$ and $\hat{D}$. In particular, we will write the multi-string interaction as
\begin{equation}\label{SemiClassicalScattering}
|C_I \rangle |D_J\rangle \rightarrow |\C_{IJ}\rangle |\D_{JI}\rangle
\end{equation}
Following the arguments of Section \ref{DetectorKinematicsSection}, this form holds will hold for elements of the semiclassical basis in the limit $l_s/r_s \rightarrow 0$. We can then treat the interaction as if our states were at fixed impact parameter, and then at large center-of-mass energy the momentum transfer is dominated by a saddle point $q_\perp \gg \Delta q_\perp$. We do not see any distortion in the shape of the wavepackets, nor do we see superpositions. The states $|\hat{C}_{IJ}\rangle, |\D_{JI}\rangle$ will thus be multi-string states with the same energies and transverse positions as the original states, but with different transverse momenta.

As in Section \ref{MultiString}, the interaction is dominated by pairwise interactions between each string of $C$ and each string of $D$. In the limit where $N_C, N_D \gg 1$ and the center-of-mass energy for each pair of strings is large, each string will suffer a large number of random, (nearly) classical deflections. Furthermore, as we chose the kinematics of each string to be independently chosen from a distribution, each of these deflections will be independent, and the central limit theorem suggests a Gaussian distribution in the momentum transfer just as in Section \ref{MultiString}.

As we will see soon, the upshot is that the transverse momenta of the smaller subsystem get randomized by the large number of strong deflections, and this will give us Equation \ref{MutualInfoResult}.

\subsection{Computing the Mutual Information}\label{PrelimResults}

We will start our computation with a preliminary result, bringing the results of Section \ref{DetectorResolutionSection} to a more useful form by expressing them in terms of mutual information.

Let us start working in the limit $S_C \gg S_D$, and make a projective measurement on $A$. This will destroy the entanglement between $D$ and $A$, and the combined system $DA$ will now be in a state
\begin{equation}
|D_J\rangle |A_J\rangle
\end{equation}
As the reference systems $R, A$ do not participate in the dynamics of $C, D$, we can forget about $A$. The detector, which we will denote $D_J$ for now, will interact with $C$ and we will obtain a state
\begin{equation}
\sum_I e^{-S_C/2}|\C_{IJ}\rangle |\D_{JI}\rangle |R_I\rangle
\end{equation}
As we chose $C$ to be in a state where the transverse kinematics of each string are independent, then just as in Section \ref{DetectorResolutionSection} each string in $D_J$ experiences a large number of independent deflections, so the total deflection will be chosen from a Gaussian distribution with spread given by Equation \ref{RMStransfer}. Due to $S_C \gg S_D$, which will generally require $N_C \gg N_D$, we will assume that there is enough kinematic freedom in $C$ so that the deflections suffered by every string in $D_J$ are independent. If we had $S_D \gg S_C$ instead, then this assumption wouldn't hold any longer: there would be constraints that relate the deflections of the various constituent strings of $D$.

This assumption of independence means that the states $|\D_{JI}\rangle$ are related to the original $|D_J\rangle$ by applying independent, Gaussian deflections to each string. The Gaussianity of these deflections is guaranteed by the central limit theorem. The density matrix
\begin{equation}
\rho_{\D_J} = \sum_I e^{-S_C} |\D_{JI}\rangle\langle \D_{JI}|
\end{equation}
then describes a mixed state with a Gaussian distribution in the transverse momenta. Thanks to the semiclassical basis, which is preserved by the $C+D$ interaction thanks to Equation \ref{SemiClassicalScattering}, we can now calculate the entropy of $\rho_{\D_J}$ from the phase-space volume.

If we only care about the leading part of the entropy $S_{\D_J}$ in the $l_s/r_s \rightarrow 0$ limit, then as in Section \ref{MultiString} we simply need to compute the logarithm of the phase-space volume occupied by the states $|D_{IJ}\rangle$ (with fixed $J$ and varied $I$), which is
\begin{equation}
\Big( \frac{(\Delta q_{\perp})^2_{RMS}}{q_{res}^2}\Big)^{N_D}
\end{equation}
and thus the entropy is given just as in Equation \ref{LogDformula} by

\begin{equation}\label{DJentropy}
S_{\D_J} \simeq 2N_D \Big( \log(\frac{r_s}{l_s}) - \log(\sqrt{N_C})\Big)
\end{equation}
We can express this in terms of the entropy $S_D$ as
\begin{equation}
S_{\D_J} \simeq \frac{1}{2}S_D
\end{equation}
With this result, let us compute the mutual information $I(\D_J : R) = S_{\D_J} + S_R - S_{\D_J R}$. We have $S_R = S_C$ by construction, and the last term can be written as $S_{\C}$, so we have

\begin{equation}
I(\D_J : R) \simeq S_{\D_J} + S_C - S_{\C}
\end{equation}

Recall now that we chose $C$ to be maximally mixed in its transverse kinematics. In our situation with elastic collisions, one can show that this implies that $\C$ must be maximally mixed in the transverse kinematics as well, and since there is no energy transfer we have $S_{\C} = S_C$. Thus we have
\begin{equation}
I(\D_J : R) = S_{\D_J} = \frac{1}{2}S_D
\end{equation}

While the above result is encouraging, and it already shows that a large amount of information can be recovered from an infaller, we can do better by using the full entangled state \ref{DAstate} instead of making a projective measurement. Still, the above result will be a useful intermediate step in the calculation of $I(\D A:R)$. As before, we have $S_{\D A R} = S_{\C} = S_C$, and thus we only need calculate $S_{\D A}$.

With $C$ and $D$ in initial states \ref{CRstate}, \ref{DAstate}, we let them interact as in Equation \ref{SemiClassicalScattering} and get a state
\begin{equation}
\sum_{I,J} e^{-S_C/2}e^{-S_D/2} |\C_{IJ}\rangle |\D_{JI}\rangle |R_I\rangle |A_J\rangle
\end{equation}
Recall that $c_I = e^{-S_C/2}, d_J = e^{-S_D/2}$ are constant since we assumed $C, D$ to be made of strings with fixed energies that are maximally mixed in their transverse kinematics. We trace out $C, R$ to obtain
\begin{equation}
\rho_{\D A} = \sum_{I, J, J'} e^{-S_C} e^{-S_D} |\D_{JI}\rangle \langle \D_{J'I}| \otimes |A_J\rangle\langle A_{J'}| \langle \C_{IJ}|\C_{IJ'}\rangle 
\end{equation}
In order to analyze this expression, it will be convenient to decompose the Hilbert spaces of $\C, \D$ into a tensor product of position and momentum labels. For example, a semiclassical wavepacket with central position $x_\perp$ and momentum $q_\perp$ will be labeled as the vector $(x_\perp, q_\perp)$. For multi-particle states, we will use the notation $I = (x_I, q_I)$ in order to denote the collection of transverse positions $x_I$ and momenta $q_I$. Recall that the $C+D$ interaction is only dependent on the positions of $C, D$, while it only immediately affects their momenta (up to corrections that are suppressed by $g_s^2$). Of course, the change in momentum can affect the detector's position at later times, but it will be convenient to only consider the state of $D$ immediately after the interaction with $C$, when its position remains unchanged compared to right before the interaction.

For $S_C \gg S_D$, we expect the states $|\C_{IJ}\rangle$ with fixed $I$ and varying $J$ to satisfy the orthogonality condition
\begin{equation}\label{OrthogonalityCondition}
\langle C_{IJ}|C_{IJ'}\rangle = \delta_{x_J, x_{J'}}
\end{equation}
Given the regime we are considering where $N_D\ll N_C$, it would be at best finely tuned for a different state of the D strings (indexed by different values of $J$) kick the strings in $C_I$ exactly the same way.  

Note that strong interactions are essential for this statement. If we had a perturbatively weak interaction, then at leading order we would have $|\C_{IJ}\rangle \simeq |C_I\rangle$, and thus we would instead have $\langle \C_{IJ} | \C_{IJ'}\rangle \simeq 1$ regardless of $J, J'$. It is only because of the eikonal resummation and the very large center-of-mass energy that we can use semi-classical interactions, and it is because of the large number of particles that we can use statistical arguments and assume that our interactions yield Gaussianly distributed momentum transfers.

Armed with the above orthogonality condition, we will now calculate the mutual information 
\begin{equation}
I(\D A: R) = S_{\D A} + S_R - S_{\D A R} = S_{\D A}
\end{equation}
Let's start by writing out the density matrix for $\D A$
\begin{equation}\label{DensityMatrixDA}
\rho_{\D A} = \sum_{I, J, J'} e^{-S_C}e^{-S_D} |\D_{J,I}\rangle \langle \D_{J',I}| \otimes |A_J\rangle\langle A_{J'}| \delta_{x_J, x_{J'}}
\end{equation}
where we obtained the delta function from Equation \ref{OrthogonalityCondition}. It will be now convenient to explicitly label each state $|D_J\rangle$ by the central transverse positions and momenta $(x_J, q_J)$ of its constituents. The $C+D$ interaction is only weakly dependent on $q_J$, it only depends on the positions $x_I, x_J$. Furthermore, its only effect on the state $|D_J\rangle = |x_J\rangle |q_J\rangle$ is a momentum change $q_J \rightarrow q_J + \delta q(x_I, x_J)$. The deflection $\delta q(x_I, x_J)$ is large and sensitively dependent on $x_J$. In Section \ref{MultiString}, we found that as we vary $x_J$, then for $S_C \gg S_D$ the momenta $q_J + \delta q(x_I, x_J)$ cover almost all available phase-space (in the sense that the logarithm of the covered phase-space is maximal).

We will thus assume that $q_J + \delta q(x_I, x_J)$ is uniformly distributed over all available transverse momenta (limited by the fixed energies of the strings). With this assumption, we can explicitly perform the sum over all $I$ in Equation \ref{DensityMatrixDA}. First, note that in the $(x_J, q_J)$ basis we can write 
\begin{equation}
|\D_{JI}\rangle\langle \D_{J'I}| = |x_J \rangle \langle x_J| \otimes |q_J + \delta q(x_I, x_J)\rangle \langle q_{J'} + \delta q(x_I, x_J)|
\end{equation}
where we used the delta-function in Equation \ref{DensityMatrixDA} to write $x_J = x_{J'}$. We can now fix $x_J, q_J$ and sum over all $I$. Due to the uniformity of $\delta q(x_I, x_J)$, the momentum $q = q_J + \delta q(x_I, x_J)$ will be uniformly distributed as well and we can write the sum as
\begin{equation}
\sum_I e^{-S_C} |\D_{JI}\rangle\langle \D_{J'I}| = |x_J\rangle \langle x_J| \otimes \sum_q e^{-S_D/2} |q\rangle \langle q + q_{J'} - q_J|
\end{equation}
In short: the states $|\D_{JI}\rangle$ have uniformly distributed transverse momenta, but because of the constraint $x_J = x_{J'}$ (from Equation \ref{DensityMatrixDA}) the states $|\D_{JI}\rangle, |\D_{J'I}\rangle$ have undergone the same deflection and thus the difference of their momenta is always $q_{J'} - q_J$ regardless of $I$. 

We can now write explicit expression for $\rho_{\D A}$. Let us decompose the reference system states as $|A_J\rangle = |A_{x_J}\rangle |A_{q_J}\rangle$, and then we write
\begin{equation}
\rho_{D A} =  \sum_{x_J} e^{-S_D/2}|x_J\rangle \langle x_J| \otimes |A_{x_J}\rangle \langle A_{x_J}| \otimes \sum_{q, q_J, q_{J'}}e^{-S_D} |q\rangle \langle q+q_{J'} - q_J|\otimes |A_{q_J}\rangle \langle A_{q_{J'}}|
\end{equation}
While the momentum part of this expression may appear complicated, its von Neumann entropy can be explicitly evaluated to be $\frac{1}{2}S_D$. In the basis $|q\rangle |A_{q_J}\rangle$, this is written as a relatively simple $e^{S_D} \times e^{S_D}$ matrix with zeros everywhere except a series of $e^{S_D/2}$ evenly spaced diagonals with entries $e^{-S_D}$. It is easy to explicitly diagonalize such matrices, we find that it has $e^{S_D/2}$ non-zero eigenvalues that are all $e^{-S_D/2}$ and thus the von Neumann entropy is $S_D/2$.

The position part of the expression is even simpler, it is a diagonal matrix with entropy $S_D/2$. Using the property of the von Neumann entropy $S(\rho_1\otimes \rho_2) = S(\rho_1) + S(\rho_2)$ we then obtain
\begin{equation}
I(\D A:R) = S(\rho_{\D A}) = S_D
\end{equation}
in this particular detector setup, up to terms that are subleading in the limit $l_s/r_s \rightarrow 0$.  

In the above calculation, we had assumed that $S_C \gg S_D$. If we were to take the opposite limit $S_D \gg S_C$, for the most easily detectable signal that we have focused on in this work, we could apply entirely analogous reasoning by exchanging the roles of $C, D$. We repeat our calculations in the same manner, rewriting $S_{\D A} = S_{\C R}$, and then we obtain $I(\D A : R) = S_C$ instead. Putting the two results together, we see that as long as $S_C, S_D$ are of significantly different magnitude, we have
\begin{equation}
I(\D A:R) = \min \lbrace S_C, S_D \rbrace
\end{equation}
in the limit where we neglect transverse momentum dependence in the amplitude.
Using semiclassical arguments, already at this level we have concretely demonstrated that an infaller's state can be (partially) recovered via longitudinal string spreading.

This assumes that there is no additional effect on the detector that destroys it or its capacity.  In this analysis, we did not analyze potential interactions of $D$ with Unruh radiation.  In our setup, $D$ is only mildly accelerating, but in principle it might interact via spreading with very near-horizon centered Unruh modes, analogously to how it interacts with $C$.  We do not anticipate that this destroys the information recovery we have obtained in our main calculation, but it is worth mentioning as a potential foreground effect.  In any case, if that effect {\it were} very large, it would be very interesting and surprising as a stringy effect in itself.

When $S_C \gg S_D$, the detector's ``capacity" is determined by the number of distinguishable transverse momentum states, which act as a recording device for the state of $C$. In the opposite limit when $S_D \gg S_C$, we still find that via the most easily detected kinematic variables we cannot fully recover the infaller's state, but we can only recover information that was encoded in transverse positions. This is to be expected: the $C+D$ interaction is only weakly dependent transverse momenta of $C$, and thus there is not a strong signal that would allow us to distinguish two states of $C$ with same positions and different momenta. This would require us to incorporate subleading corrections (e.g. account for the contribution of the transverse momenta to the center-of-mass energy $s_{CD}$). That may be accessible in the large $N_D$ case via effectively repeated independent experiments, but $N_D$ itself is limited by back reaction constraints. 

It is worth stressing that these limitations apply to the particular setup of semiclassical product wavefunctions we worked with here. An optimal detector setup would also generalize our system to incorporate entanglement among the constituents of $D$.   Moreover, it is possible that effects that lead to particle production (e.g. stringy inelastic scattering, microscopic black hole formation) can lead to more universal information recovery.   Generalizing the detector system to incorporate D-branes would also be interesting.  Regardless, we found substantial information transfer in our simpler setup of semiclassical $C$-$D$ interaction.

%\section{Discussion}
%
%Future directions:
%
%\begin{itemize}
%
%\item Extend to string-brane scattering: If a brane could absorb a closed string via longitudinal spreading, it would provide an easy way to recover information, especially information encoded in quantum numbers. The way to study this would be to compute 5-point closed-open string amplitudes on a disk.
%
%
%\item Spreading in gauge theories: There should be some finite $\lambda$ effect that yields spreading from a boundary perspective. Since the bulk mechanism is connected to gravitational effects, perhaps one could find finite $\lambda$ corrections to the Virasoro conformal block that achieve a similar effect.
%
%\end{itemize}
%

\section*{Acknowledgements}

We would like to thank R. Bousso, A. Brown,  M. Dodelson, P. Hayden, D. Mathis, and D. Stanford for useful discussions.  
This research is supported by the  National Science Foundation under grant number PHY-1720397, a Simons Foundation Investigator Award,  and a Stanford Graduate Fellowship.  

\newpage

\appendix

\begingroup\raggedright\begin{thebibliography}{10}
\baselineskip=14.5pt

\bibitem{lennyspreading}
L. Susskind,
``Strings, black holes and Lorentz contraction,"
Phys. Rev. D {\bf 49}, 6606-6611 (1994). \\ 
M. Karliner, I. R. Klebanov, and L. Susskind,
``Size and Shape of Strings," 
Int. J. Mod. Phys. {\bf A3} 1981 (1988). 
[hep-th/9308139].

\bibitem{BHpaper}
%\bibitem{Dodelson:2015toa} 
  M.~Dodelson and E.~Silverstein,
  ``String-theoretic breakdown of effective field theory near black hole horizons,''
  arXiv:1504.05536 [hep-th].
  %%CITATION = ARXIV:1504.05536;%%
  %1 citations counted in INSPIRE as of 20 May 2015

\bibitem{sixpoints} 
  M.~Dodelson and E.~Silverstein,
  ``Long-Range Nonlocality in Six-Point String Scattering: simulation of black hole infallers,''
  arXiv:1703.10147 [hep-th].
  %%CITATION = ARXIV:1703.10147;%%
  %1 citations counted in INSPIRE as of 01 Jun 2017

\bibitem{dilatontracer}

M.~Dodelson, E.~Silverstein and G.~Torroba,
  ``Varying dilaton as a tracer of classical string interactions,''
  arXiv:1704.02625 [hep-th].
  %%CITATION = ARXIV:1704.02625;%%
%\cite{Dodelson:2017hyu}
  
\bibitem{Backdraft}

A.~Puhm, F.~Rojas and T.~Ugajin,
  ``(Non-adiabatic) string creation on nice slices in Schwarzschild black holes,''
  JHEP {\bf 1704}, 156 (2017)
  doi:10.1007/JHEP04(2017)156
  [arXiv:1609.09510 [hep-th]].
  %%CITATION = doi:10.1007/JHEP04(2017)156;%%
  %3 citations counted in INSPIRE as of 01 Jun 2017

 E.~Silverstein,
  ``Backdraft: String Creation in an Old Schwarzschild Black Hole,''
  arXiv:1402.1486 [hep-th].
  %%CITATION = ARXIV:1402.1486;%%
  %26 citations counted in INSPIRE as of 01 Jun 2017

\bibitem{Danjie}

D.~Wenren,
  ``Hyperbolic Black Holes and Open String Production,''
  arXiv:1709.03590 [hep-th].
  %%CITATION = ARXIV:1709.03590;%%

\bibitem{grossmende}
 D.~J.~Gross and P.~F.~Mende,
  ``String Theory Beyond the Planck Scale,''
  Nucl.\ Phys.\ B {\bf 303}, 407 (1988).
  %%CITATION = NUPHA,B303,407;%%
  %776 citations counted in INSPIRE as of 11 Aug 2015
  
P.~F.~Mende and H.~Ooguri,
  ``Borel Summation of String Theory for Planck Scale Scattering,''
  Nucl.\ Phys.\ B {\bf 339}, 641 (1990).
  doi:10.1016/0550-3213(90)90202-O
  %%CITATION = doi:10.1016/0550-3213(90)90202-O;%%
  %56 citations counted in INSPIRE as of 05 Oct 2018  

\bibitem{Warpedpaper}

M. Dodelson, B. Kang, D. Mathis, A. Mousatov, E. Silverstein,  in preparation

\bibitem{JoeReview}

J.~Polchinski,
  ``The Black Hole Information Problem,''
  %$doi:10.1142/9789813149441_0006$
  arXiv:1609.04036 [hep-th].
  %%CITATION = doi:10.1142/9789813149441_0006;%%
  %54 citations counted in INSPIRE as of 29 Nov 2019

\bibitem{PageCurveW}

%\bibitem{Penington:2019kki} 
  G.~Penington, S.~H.~Shenker, D.~Stanford and Z.~Yang,
  ``Replica wormholes and the black hole interior,''
  arXiv:1911.11977 [hep-th].
  %%CITATION = ARXIV:1911.11977;%%
  %1 citations counted in INSPIRE as of 04 Dec 2019

\bibitem{PageCurveE}

%\cite{Almheiri:2019qdq}
%\bibitem{Almheiri:2019qdq} 
  A.~Almheiri, T.~Hartman, J.~Maldacena, E.~Shaghoulian and A.~Tajdini,
  ``Replica Wormholes and the Entropy of Hawking Radiation,''
  arXiv:1911.12333 [hep-th].
  %%CITATION = ARXIV:1911.12333;%%
  %1 citations counted in INSPIRE as of 04 Dec 2019
%\cite{Penington:2019kki}

\bibitem{EWreconstruction}

C.~Akers, A.~Levine and S.~Leichenauer,
  ``Large Breakdowns of Entanglement Wedge Reconstruction,''
  arXiv:1908.03975 [hep-th].
  %%CITATION = ARXIV:1908.03975;%%
  %5 citations counted in INSPIRE as of 04 Dec 2019
  
  A.~Lewkowycz and O.~Parrikar,
  ``The holographic shape of entanglement and Einstein’s equations,''
  JHEP {\bf 1805}, 147 (2018)
  doi:10.1007/JHEP05(2018)147
  [arXiv:1802.10103 [hep-th]].
  %%CITATION = doi:10.1007/JHEP05(2018)147;%%
  %14 citations counted in INSPIRE as of 04 Dec 2019
  
  T.~Faulkner and A.~Lewkowycz,
  ``Bulk locality from modular flow,''
  JHEP {\bf 1707}, 151 (2017)
  doi:10.1007/JHEP07(2017)151
  [arXiv:1704.05464 [hep-th]].
  %%CITATION = doi:10.1007/JHEP07(2017)151;%%
  %80 citations counted in INSPIRE as of 04 Dec 2019
  
   X.~Dong, A.~Lewkowycz and M.~Rangamani,
  ``Deriving covariant holographic entanglement,''
  JHEP {\bf 1611}, 028 (2016)
  doi:10.1007/JHEP11(2016)028
  [arXiv:1607.07506 [hep-th]].
  %%CITATION = doi:10.1007/JHEP11(2016)028;%%
  %105 citations counted in INSPIRE as of 04 Dec 2019
  
  D.~L.~Jafferis, A.~Lewkowycz, J.~Maldacena and S.~J.~Suh,
  ``Relative entropy equals bulk relative entropy,''
  JHEP {\bf 1606}, 004 (2016)
  doi:10.1007/JHEP06(2016)004
  [arXiv:1512.06431 [hep-th]].
  %%CITATION = doi:10.1007/JHEP06(2016)004;%%
  %176 citations counted in INSPIRE as of 04 Dec 2019

  T.~Faulkner, A.~Lewkowycz and J.~Maldacena,
  ``Quantum corrections to holographic entanglement entropy,''
  JHEP {\bf 1311}, 074 (2013)
  doi:10.1007/JHEP11(2013)074
  [arXiv:1307.2892 [hep-th]].
  %%CITATION = doi:10.1007/JHEP11(2013)074;%%
  %320 citations counted in INSPIRE as of 04 Dec 2019
%\cite{Lewkowycz:2013nqa}

X.~Dong, D.~Harlow and A.~C.~Wall,
  ``Reconstruction of Bulk Operators within the Entanglement Wedge in Gauge-Gravity Duality,''
  Phys.\ Rev.\ Lett.\  {\bf 117}, no. 2, 021601 (2016)
  doi:10.1103/PhysRevLett.117.021601
  [arXiv:1601.05416 [hep-th]].
  %%CITATION = doi:10.1103/PhysRevLett.117.021601;%%
  %173 citations counted in INSPIRE as of 04 Dec 2019

M.~Headrick, V.~E.~Hubeny, A.~Lawrence and M.~Rangamani,
  ``Causality $\&$ holographic entanglement entropy,''
  JHEP {\bf 1412}, 162 (2014)
  doi:10.1007/JHEP12(2014)162
  [arXiv:1408.6300 [hep-th]].
  %%CITATION = doi:10.1007/JHEP12(2014)162;%%
  %174 citations counted in INSPIRE as of 04 Dec 2019

%\bibitem{Lewkowycz:2013nqa} 
  A.~Lewkowycz and J.~Maldacena,
  ``Generalized gravitational entropy,''
  JHEP {\bf 1308}, 090 (2013)
  doi:10.1007/JHEP08(2013)090
  [arXiv:1304.4926 [hep-th]].
  %%CITATION = doi:10.1007/JHEP08(2013)090;%%
  %523 citations counted in INSPIRE as of 04 Dec 2019

\bibitem{Islands}

G.~Penington,
  ``Entanglement Wedge Reconstruction and the Information Paradox,''
  arXiv:1905.08255 [hep-th].
  %%CITATION = ARXIV:1905.08255;%%
  %25 citations counted in INSPIRE as of 04 Dec 2019

%\cite{Almheiri:2019psf}
%\bibitem{Almheiri:2019psf} 
  A.~Almheiri, N.~Engelhardt, D.~Marolf and H.~Maxfield,
  ``The entropy of bulk quantum fields and the entanglement wedge of an evaporating black hole,''
  arXiv:1905.08762 [hep-th].
  %%CITATION = ARXIV:1905.08762;%%
  %24 citations counted in INSPIRE as of 04 Dec 2019

%\cite{Almheiri:2019yqk}
%\bibitem{Almheiri:2019yqk} 
  A.~Almheiri, R.~Mahajan and J.~Maldacena,
  ``Islands outside the horizon,''
  arXiv:1910.11077 [hep-th].
  %%CITATION = ARXIV:1910.11077;%%

%\cite{Almheiri:2019psy}
%\bibitem{Almheiri:2019psy} 
  A.~Almheiri, R.~Mahajan and J.~E.~Santos,
  ``Entanglement islands in higher dimensions,''
  arXiv:1911.09666 [hep-th].
  %%CITATION = ARXIV:1911.09666;%%
  %2 citations counted in INSPIRE as of 04 Dec 2019
  
\bibitem{MarolfMaxfield}

 D.~Marolf and H.~Maxfield,
  ``Transcending the ensemble: baby universes, spacetime wormholes, and the order and disorder of black hole information,''
  arXiv:2002.08950 [hep-th].
  %%CITATION = ARXIV:2002.08950;%%

\bibitem{InfoCommentary}

%\cite{Bousso:2019ykv}
%\bibitem{Bousso:2019ykv} 
  R.~Bousso and M.~Tomasevic,
  ``Unitarity From a Smooth Horizon?,''
  arXiv:1911.06305 [hep-th].
  %%CITATION = ARXIV:1911.06305;%%

C.~Akers, N.~Engelhardt and D.~Harlow,
  ``Simple holographic models of black hole evaporation,''
  arXiv:1910.00972 [hep-th].
  %%CITATION = ARXIV:1910.00972;%%
  %7 citations counted in INSPIRE as of 04 Dec 2019
  
  A.~R.~Brown, H.~Gharibyan, G.~Penington and L.~Susskind,
  ``The Python's Lunch: geometric obstructions to decoding Hawking radiation,''
  arXiv:1912.00228 [hep-th].
  %%CITATION = ARXIV:1912.00228;%%

\bibitem{TTEE}

A.~Lewkowycz, J.~Liu, E.~Silverstein and G.~Torroba,
  ``$T \bar T$ and EE, with implications for (A)dS subregion encodings,''
  arXiv:1909.13808 [hep-th].
  %%CITATION = ARXIV:1909.13808;%%
  %5 citations counted in INSPIRE as of 04 Dec 2019
  
  C.~Murdia, Y.~Nomura, P.~Rath and N.~Salzetta,
  ``Comments on holographic entanglement entropy in $TT$ deformed conformal field theories,''
  Phys.\ Rev.\ D {\bf 100}, no. 2, 026011 (2019)
  doi:10.1103/PhysRevD.100.026011
  [arXiv:1904.04408 [hep-th]].
  %%CITATION = doi:10.1103/PhysRevD.100.026011;%%
  %10 citations counted in INSPIRE as of 04 Dec 2019

\bibitem{JuanST}

X.~O.~Camanho, J.~D.~Edelstein, J.~Maldacena and A.~Zhiboedov,
  ``Causality Constraints on Corrections to the Graviton Three-Point Coupling,''
  JHEP {\bf 1602}, 020 (2016)
  doi:10.1007/JHEP02(2016)020
  [arXiv:1407.5597 [hep-th]].
  %%CITATION = doi:10.1007/JHEP02(2016)020;%%
  %304 citations counted in INSPIRE as of 04 Dec 2019
  
\bibitem{STtimelike}  

L.~J.~Dixon, J.~A.~Harvey, C.~Vafa and E.~Witten,
  ``Strings on Orbifolds,''
  Nucl.\ Phys.\ B {\bf 261}, 678 (1985).
  doi:10.1016/0550-3213(85)90593-0
  %%CITATION = doi:10.1016/0550-3213(85)90593-0;%%
  %1622 citations counted in INSPIRE as of 04 Dec 2019

A.~Adams, J.~Polchinski and E.~Silverstein,
  ``Don't panic! Closed string tachyons in ALE space-times,''
  JHEP {\bf 0110}, 029 (2001)
  doi:10.1088/1126-6708/2001/10/029
  [hep-th/0108075].
  %%CITATION = doi:10.1088/1126-6708/2001/10/029;%%
  %272 citations counted in INSPIRE as of 04 Dec 2019

E.~Witten,
  ``Phases of N=2 theories in two-dimensions,''
  Nucl.\ Phys.\ B {\bf 403}, 159 (1993)
  [AMS/IP Stud.\ Adv.\ Math.\  {\bf 1}, 143 (1996)]
  doi:10.1016/0550-3213(93)90033-L
  [hep-th/9301042].
  %%CITATION = doi:10.1016/0550-3213(93)90033-L;%%
  %1233 citations counted in INSPIRE as of 04 Dec 2019  
  
P.~S.~Aspinwall, B.~R.~Greene and D.~R.~Morrison,
  ``Multiple mirror manifolds and topology change in string theory,''
  Phys.\ Lett.\ B {\bf 303}, 249 (1993)
  doi:10.1016/0370-2693(93)91428-P
  [hep-th/9301043].
  %%CITATION = doi:10.1016/0370-2693(93)91428-P;%%
  %112 citations counted in INSPIRE as of 04 Dec 2019  
  
 P.~S.~Aspinwall, B.~R.~Greene and D.~R.~Morrison,
  ``Calabi-Yau moduli space, mirror manifolds and space-time topology change in string theory,''
  Nucl.\ Phys.\ B {\bf 416}, 414 (1994)
  [AMS/IP Stud.\ Adv.\ Math.\  {\bf 1}, 213 (1996)]
  doi:10.1016/0550-3213(94)90321-2
  [hep-th/9309097].
  %%CITATION = doi:10.1016/0550-3213(94)90321-2;%%
  %285 citations counted in INSPIRE as of 04 Dec 2019 
  
J.~Distler and S.~Kachru,
  ``Duality of (0,2) string vacua,''
  Nucl.\ Phys.\ B {\bf 442}, 64 (1995)
  doi:10.1016/S0550-3213(95)00130-1
  [hep-th/9501111].
  %%CITATION = doi:10.1016/S0550-3213(95)00130-1;%%
  %47 citations counted in INSPIRE as of 12 Feb 2020  
  
A.~Adams, X.~Liu, J.~McGreevy, A.~Saltman and E.~Silverstein,
  ``Things fall apart: Topology change from winding tachyons,''
  JHEP {\bf 0510}, 033 (2005)
  doi:10.1088/1126-6708/2005/10/033
  [hep-th/0502021].
  %%CITATION = doi:10.1088/1126-6708/2005/10/033;%%
  %69 citations counted in INSPIRE as of 04 Dec 2019
  
\bibitem{Plancktimelike}

A.~Strominger,
  ``Massless black holes and conifolds in string theory,''
  Nucl.\ Phys.\ B {\bf 451}, 96 (1995)
  doi:10.1016/0550-3213(95)00287-3
  [hep-th/9504090].
  %%CITATION = doi:10.1016/0550-3213(95)00287-3;%%
  %628 citations counted in INSPIRE as of 04 Dec 2019
  
C.~V.~Johnson, A.~W.~Peet and J.~Polchinski,
  ``Gauge theory and the excision of repulson singularities,''
  Phys.\ Rev.\ D {\bf 61}, 086001 (2000)
  doi:10.1103/PhysRevD.61.086001
  [hep-th/9911161].
  %%CITATION = doi:10.1103/PhysRevD.61.086001;%%
  %255 citations counted in INSPIRE as of 04 Dec 2019

\bibitem{STspacelike}

 G.~T.~Horowitz,
  ``Tachyon condensation and black strings,''
  JHEP {\bf 0508}, 091 (2005)
  doi:10.1088/1126-6708/2005/08/091
  [hep-th/0506166].
  %%CITATION = doi:10.1088/1126-6708/2005/08/091;%%
  %76 citations counted in INSPIRE as of 19 Feb 2020

G.~T.~Horowitz and E.~Silverstein,
  ``The Inside story: Quasilocal tachyons and black holes,''
  Phys.\ Rev.\ D {\bf 73}, 064016 (2006)
  doi:10.1103/PhysRevD.73.064016
  [hep-th/0601032].
  %%CITATION = doi:10.1103/PhysRevD.73.064016;%%
  %67 citations counted in INSPIRE as of 04 Dec 2019  
  
 J.~McGreevy and E.~Silverstein,
  ``The Tachyon at the end of the universe,''
  JHEP {\bf 0508}, 090 (2005)
  doi:10.1088/1126-6708/2005/08/090
  [hep-th/0506130].
  %%CITATION = doi:10.1088/1126-6708/2005/08/090;%%
  %110 citations counted in INSPIRE as of 04 Dec 2019 

\bibitem{Wittenbubble}

 E.~Witten,
  ``Instability of the Kaluza-Klein Vacuum,''
  Nucl.\ Phys.\ B {\bf 195}, 481 (1982).
  doi:10.1016/0550-3213(82)90007-4
  %%CITATION = doi:10.1016/0550-3213(82)90007-4;%%
  %443 citations counted in INSPIRE as of 12 Feb 2020
  
\bibitem{correspondence}

G.~T.~Horowitz and J.~Polchinski,
  ``A Correspondence principle for black holes and strings,''
  Phys.\ Rev.\ D {\bf 55}, 6189 (1997)
  doi:10.1103/PhysRevD.55.6189
  [hep-th/9612146].
  %%CITATION = doi:10.1103/PhysRevD.55.6189;%%
  %536 citations counted in INSPIRE as of 12 Feb 2020  
  
L.~Susskind,
  ``Some speculations about black hole entropy in string theory,''
  In *Teitelboim, C. (ed.): The black hole* 118-131
  [hep-th/9309145].
  %%CITATION = HEP-TH/9309145;%%
  %475 citations counted in INSPIRE as of 12 Feb 2020  

\bibitem{ACV}

D.~Amati, M.~Ciafaloni and G.~Veneziano,
  ``Superstring Collisions at Planckian Energies,''
  Phys.\ Lett.\ B {\bf 197}, 81 (1987).
  %%CITATION = PHLTA,B\hat{B}97,81;%%
  %450 citations counted in INSPIRE as of 31 Aug 2015

D.~Amati, M.~Ciafaloni and G.~Veneziano,
  ``Can Space-Time Be Probed Below the String Size?,''
  Phys.\ Lett.\ B {\bf 216}, 41 (1989).
  %%CITATION = PHLTA,B216,41;%%
  %723 citations counted in INSPIRE as of 31 Aug 2015
\bibitem{locality}
D. Lowe, J. Polchinski, L. Susskind, L. Thorlacius, J. Uglum,
``Black hole complementarity versus locality,"
Phys. Rev. D {\bf{52}} 6997-7010 (1995) [hep-th/9506138].
%\cite{Polchinski:1995ta}

\bibitem{Giddings}

S.~B.~Giddings,
  ``The gravitational S-matrix: Erice lectures,''
  Subnucl.\ Ser.\  {\bf 48}, 93 (2013)
  %doi:10.1142/9789814522489_0005
  [arXiv:1105.2036 [hep-th]].
  %%CITATION = doi:10.1142/9789814522489_0005;%%
  %50 citations counted in INSPIRE as of 29 Mar 2018

\bibitem{Weinbergsoft}

 S.~Weinberg,
  ``Infrared photons and gravitons,''
  Phys.\ Rev.\  {\bf 140}, B516 (1965).
  doi:10.1103/PhysRev.140.B516
  %%CITATION = doi:10.1103/PhysRev.140.B516;%%
  %533 citations counted in INSPIRE as of 31 May 2018  

S. Weinberg, 
{\it The Quantum Theory of Fields} volume  I

\bibitem{JoeBook}

J.~Polchinski,
  ``String theory. Vol. 1: An introduction to the bosonic string,''
  Cambridge, UK: Univ. Pr. (1998) 402 p
  
\bibitem{BPST}
R. C. Brower, J. Polchinski, M. J. Strassler, C. Tan,
``The Pomeron and gauge/string duality,"
JHEP {\bf{0712}} 005 (2007) [hep-th/0603115].
  
\bibitem{MuzinichSoldate}

I.~J.~Muzinich and M.~Soldate,
  ``High-Energy Unitarity of Gravitation and Strings,''
  Phys.\ Rev.\ D {\bf 37}, 359 (1988).
  doi:10.1103/PhysRevD.37.359
  %%CITATION = doi:10.1103/PhysRevD.37.359;%%
  %132 citations counted in INSPIRE as of 31 Mar 2018

\bibitem{Kabat}

D.~N.~Kabat and M.~Ortiz,
  ``Eikonal quantum gravity and Planckian scattering,''
  Nucl.\ Phys.\ B {\bf 388}, 570 (1992)
  doi:10.1016/0550-3213(92)90627-N
  [hep-th/9203082].
  %%CITATION = doi:10.1016/0550-3213(92)90627-N;%%
  %101 citations counted in INSPIRE as of 31 Mar 2018
  
\bibitem{iepsilon}

 A.~Berera,
  ``Unitary string amplitudes,''
  Nucl.\ Phys.\ B {\bf 411}, 157 (1994).
  doi:10.1016/0550-3213(94)90057-4
  %%CITATION = doi:10.1016/0550-3213(94)90057-4;%%
  %26 citations counted in INSPIRE as of 02 Apr 2018    

E.~Witten,
  ``The Feynman $i \epsilon$ in String Theory,''
  JHEP {\bf 1504}, 055 (2015)
  doi:10.1007/JHEP04(2015)055
  [arXiv:1307.5124 [hep-th]].
  %%CITATION = doi:10.1007/JHEP04(2015)055;%%
  %34 citations counted in INSPIRE as of 02 Apr 2018

\bibitem{JoeBHcomp} 
  J.~Polchinski,
  ``String theory and black hole complementarity,''
  In *Los Angeles 1995, Future perspectives in string theory* 417-426
  [hep-th/9507094].
  %%CITATION = HEP-TH/9507094;%%
  %28 citations counted in INSPIRE as of 04 Feb 2015
  
  \bibitem{ggm}
S.~B.~Giddings, D.~J.~Gross and A.~Maharana,
  ``Gravitational effects in ultrahigh-energy string scattering,''
  Phys.\ Rev.\ D {\bf 77}, 046001 (2008)
  [arXiv:0705.1816 [hep-th]].
  %%CITATION = ARXIV:0705.1816;%%
  %55 citations counted in INSPIRE as of 31 Aug 2015

\bibitem{oldmining}

W.~G.~Unruh and R.~M.~Wald,
  ``Acceleration Radiation and Generalized Second Law of Thermodynamics,''
  Phys.\ Rev.\ D {\bf 25}, 942 (1982).
  doi:10.1103/PhysRevD.25.942
  %%CITATION = doi:10.1103/PhysRevD.25.942;%%
  %255 citations counted in INSPIRE as of 05 Oct 2018

Unruh, W.G. and Wald, R.~M. ``How to mine energy from a black hole'', R.M. Gen Relat Gravit (1983) 15: 195. https://doi.org/10.1007/BF00759206

\bibitem{AdamMining}

A.~R.~Brown,
  ``Tensile Strength and the Mining of Black Holes,''
  Phys.\ Rev.\ Lett.\  {\bf 111}, no. 21, 211301 (2013)
  doi:10.1103/PhysRevLett.111.211301
  [arXiv:1207.3342 [gr-qc]].
  %%CITATION = doi:10.1103/PhysRevLett.111.211301;%%
  %30 citations counted in INSPIRE as of 17 Mar 2018

\bibitem{AMPS}

 A.~Almheiri, D.~Marolf, J.~Polchinski and J.~Sully,
  ``Black Holes: Complementarity or Firewalls?,''
  JHEP {\bf 1302}, 062 (2013)
  doi:10.1007/JHEP02(2013)062
  [arXiv:1207.3123 [hep-th]].
  %%CITATION = doi:10.1007/JHEP02(2013)062;%%
  %601 citations counted in INSPIRE as of 09 Jun 2017

\bibitem{complementarity}

L.~Susskind, L.~Thorlacius and J.~Uglum,
  ``The Stretched horizon and black hole complementarity,''
  Phys.\ Rev.\ D {\bf 48}, 3743 (1993)
  doi:10.1103/PhysRevD.48.3743
  [hep-th/9306069].
  %%CITATION = doi:10.1103/PhysRevD.48.3743;%%
  %726 citations counted in INSPIRE as of 09 Jun 2017

\bibitem{HaydenPreskill}

P. Hayden, J. Preskill,
	``Black holes as mirrors: Quantum information in random subsystems,''
	JHEP 0709 (2007) 120, 
	doi:10.1088/1126-6708/2007/09/120
	[arXiv:0708.4025 [hep-th]]

%\cite{Hawking:1974sw}
\bibitem{HawkingRadiation} 
  S.~W.~Hawking,
  ``Particle Creation by Black Holes,''
  Commun.\ Math.\ Phys.\  {\bf 43}, 199 (1975)
  Erratum: [Commun.\ Math.\ Phys.\  {\bf 46}, 206 (1976)].
  doi:10.1007/BF02345020, 10.1007/BF01608497
  %%CITATION = doi:10.1007/BF02345020, 10.1007/BF01608497;%%
  %7288 citations counted in INSPIRE as of 03 Oct 2018

\bibitem{ChaosBHSmatrix}

J.~Polchinski,
	``Chaos in the black hole S-matrix,''
	[arXiv:1505.08108 [hep-th]]

\bibitem{SSchaos}

%\cite{Shenker:2014cwa}
%\bibitem{Shenker:2014cwa} 
  S.~H.~Shenker and D.~Stanford,
  ``Stringy effects in scrambling,''
  JHEP {\bf 1505}, 132 (2015)
  doi:10.1007/JHEP05(2015)132
  [arXiv:1412.6087 [hep-th]].
  %%CITATION = doi:10.1007/JHEP05(2015)132;%%
  %169 citations counted in INSPIRE as of 04 Oct 2018

%\cite{Shenker:2013pqa}
%\bibitem{Shenker:2013pqa} 
  S.~H.~Shenker and D.~Stanford,
  ``Black holes and the butterfly effect,''
  JHEP {\bf 1403}, 067 (2014)
  doi:10.1007/JHEP03(2014)067
  [arXiv:1306.0622 [hep-th]].
  %%CITATION = doi:10.1007/JHEP03(2014)067;%%
  %373 citations counted in INSPIRE as of 04 Oct 2018

\bibitem{GSW}

 M.~B.~Green, J.~H.~Schwarz and E.~Witten,
  ``Superstring Theory. Vol. 1: Introduction,''
  Cambridge Monographs in  Mathematical Physics (1987)
  %184 citations counted in INSPIRE as of 20 Oct 2016

\bibitem{JoeBook}

J.~Polchinski,
  ``String theory. Vol. 1: An introduction to the bosonic string,''
  Cambridge, UK: Univ. Pr. (1998) 402 p

%\bibitem{strassler5pt}
%C. P. Herzog, S. Paik, M. J. Strassler, E. G. Thompson, 
%``Holographic Double Diffractive Scattering," JHEP 0808 (2008) 010 [hep-th/08060181].
%S. Giddings, private discussions

%\bibitem{sister}
%C. Barratt, ``Multi-Regge Limit of the Virasoro-Shapiro Model: A Sister for the Pomeron," Nucl. Phys. B126 (1977) 133;\\
%P. Hoyer, N. A. Tornqvist, B.R. Webber, ``A new Regge trajectory in the dual resonance model," Phys. Lett. B61 (1976) 191;

%\cite{Witten:2013pra}

\bibitem{BHreviews}

J.~Polchinski,
  ``The Black Hole Information Problem,''
  arXiv:1609.04036 [hep-th].
  %%CITATION = ARXIV:1609.04036;%%
  %2 citations counted in INSPIRE as of 04 Nov 2016

 D.~Harlow,
  ``Jerusalem Lectures on Black Holes and Quantum Information,''
  arXiv:1409.1231 [hep-th].
  %%CITATION = ARXIV:1409.1231;%%
  %23 citations counted in INSPIRE as of 17 Nov 2015

\bibitem{CorrespondenceBHstrings}

G. T. Horowitz, J. Polchinski, 
	``A Correspondence principle for black holes and strings,''
	Phys.Rev. D55 (1997) 6189-6197
	doi:10.1103/PhysRevD.55.6189
	[arxiv:9612146 [hep-th]].

\bibitem{otherstringy}

R.~Ben-Israel, A.~Giveon, N.~Itzhaki and L.~Liram,
  ``Stringy Horizons and UV/IR Mixing,''
  arXiv:1506.07323 [hep-th].
  %%CITATION = ARXIV:1506.07323;%%
  %3 citations counted in INSPIRE as of 17 Nov 2015

A.~Giveon and N.~Itzhaki,
  ``String Theory Versus Black Hole Complementarity,''
  JHEP {\bf 1212}, 094 (2012)
  doi:10.1007/JHEP12(2012)094
  [arXiv:1208.3930 [hep-th]].
  %%CITATION = doi:10.1007/JHEP12(2012)094;%%
  %39 citations counted in INSPIRE as of 17 Nov 2015

 I.~Bena, G.~Bossard, S.~Katmadas and D.~Turton,
  ``Non-BPS multi-bubble microstate geometries,''
  arXiv:1511.03669 [hep-th].
  %%CITATION = ARXIV:1511.03669;%%
%\cite{Bena:2015lkx}

%\bibitem{Mertens:2015adr} 
  T.~G.~Mertens, H.~Verschelde and V.~I.~Zakharov,
  ``String partition functions in Rindler space and maximal acceleration,''
  arXiv:1511.00560 [hep-th].
  %%CITATION = ARXIV:1511.00560;%%

%\cite{Giddings:2015uzr}
\bibitem{HawkingLocation} 

  S.~B.~Giddings,
  ``Hawking radiation, the Stefan–Boltzmann law, and unitarization,''
  Phys.\ Lett.\ B {\bf 754}, 39 (2016)
  doi:10.1016/j.physletb.2015.12.076
  [arXiv:1511.08221 [hep-th]].
  %%CITATION = doi:10.1016/j.physletb.2015.12.076;%%
  %16 citations counted in INSPIRE as of 02 Jun 2017

\bibitem{NonviolentNonlocality}

	S.~B.~Giddings,
	``Nonviolent nonlocality,''
	Phys.Rev. D88 (2013) 064023
	doi:10.1103/PhysRevD.88.064023
	[arxiv:1211.7070 [hep-th]].

\bibitem{HodlargeD}

S.~Hod,
  ``Hawking radiation and the Stefan-Boltzmann law: The effective radius of the black-hole quantum atmosphere,''
  Phys.\ Lett.\ B {\bf 757}, 121 (2016)
  doi:10.1016/j.physletb.2016.03.071
  [arXiv:1607.02510 [gr-qc]].
  %%CITATION = doi:10.1016/j.physletb.2016.03.071;%%
  %3 citations counted in INSPIRE as of 02 Jun 2017

%\bibitem{ScatteringEntanglement}
%
%R. Peschanski, S. Seki, 
%	``Entanglement Entropy of Scattering Particles,''
%	Phys.\ Lett.\ B {\bf 758}, 89-92 (2016)
%	doi:10.1016/j.physletb.2016.04.063
%	[arxiv:arXiv:1602.00720 [hep-th]].

%\bibitem{largeDmodels}
%
%E.~Silverstein,
%  ``(A)dS backgrounds from asymmetric orientifolds,''
%  hep-th/0106209.
%  %%CITATION = HEP-TH/0106209;%%
%  %131 citations counted in INSPIRE as of 02 Jun 2017
%
%A.~Maloney, E.~Silverstein and A.~Strominger,
%  ``De Sitter space in noncritical string theory,''
%  hep-th/0205316.
%  %%CITATION = HEP-TH/0205316;%%
%  %155 citations counted in INSPIRE as of 02 Jun 2017
%
% M.~Dodelson, X.~Dong, E.~Silverstein and G.~Torroba,
%  ``New solutions with accelerated expansion in string theory,''
%  JHEP {\bf 1412}, 050 (2014)
%  doi:10.1007/JHEP12(2014)050
%  [arXiv:1310.5297 [hep-th]].
%  %%CITATION = doi:10.1007/JHEP12(2014)050;%%
%  %13 citations counted in INSPIRE as of 02 Jun 2017

  %62 citations counted in INSPIRE as of 19 Dec 2014
\endgroup
\end{document}